\documentclass[11pt,a4paper]{article}
\usepackage{jcappub}
\usepackage{amsmath,amssymb,epsf}
\usepackage{graphicx,color}
\usepackage{amsfonts}
\usepackage{hyperref}

\pdfoutput=1

\def\d{{\rm d}}

\def\unit{\relax{\rm 1\kern-.26em I}}
\def\nada{\relax{\rm 0\kern-.30em l}}

\numberwithin{equation}{section}
\newcommand{\be}{\begin{equation}}
\newcommand{\ee}{\end{equation}}
\newcommand{\bea}{\begin{eqnarray}}
\newcommand{\eea}{\end{eqnarray}}
\newcommand{\barr}{\begin{array}}
\newcommand{\earr}{\end{array}}

\def\beq{\begin{equation}}
\def\eeq{\end{equation}}
\def\be{\begin{equation}}
\def\ee{\end{equation}}
\def\bea{\begin{eqnarray}}
\def\eea{\end{eqnarray}}
\def\d{{\rm d}}
\DeclareRobustCommand{\SkipTocEntry}[4]{}

\title{Just enough inflation: power spectrum modifications at large scales}
\author[a,b,c]{Michele Cicoli,}
\author[d]{Sean Downes,}
\author[e]{Bhaskar Dutta,}
\author[f]{Francisco G. Pedro,}
\author[f]{Alexander Westphal}
\affiliation[a]{Dipartimento di Fisica ed Astronomia, Universit\`a di Bologna, \\ via Irnerio 46, 40126 Bologna, Italy}
\affiliation[b]{INFN, Sezione di Bologna, 40126 Bologna, Italy}
\affiliation[c]{Abdus Salam ICTP, Strada Costiera 11, Trieste 34014, Italy}
\affiliation[d]{Leung Center for Cosmology and Particle Astrophysics National Taiwan University}
\affiliation[e]{Mitchell Institute for Fundamental Physics and Astronomy, \\ Department of Physics and Astronomy, Texas A\&M University, \\ College Station, TX 77843-4242, USA}
\affiliation[f]{Deutsches Elektronen-Synchrotron DESY, Theory Group, D-22603 Hamburg, Germany.}
\emailAdd{mcicoli@ictp.it}
\emailAdd{ssdownes@phys.ntu.edu.tw}
\emailAdd{dutta@physics.tamu.edu}
\emailAdd{francisco.pedro@desy.de}
\emailAdd{alexander.westphal@desy.de}

\abstract{We show that models of `just enough' inflation, where the slow-roll evolution lasted only $50-60$ e-foldings,
feature modifications of the CMB power spectrum at large angular scales.
We perform a systematic and model-independent analysis of any possible non-slow-roll background evolution prior to the final stage of slow-roll inflation.
We find a high degree of universality since most common backgrounds like fast-roll evolution, matter or radiation-dominance give rise to a power loss at
large angular scales and a peak together with an oscillatory behaviour at scales around the value of the Hubble parameter at the beginning of slow-roll inflation. Depending on the value of the equation of state parameter, different pre-inflationary epochs lead instead to an enhancement of power at low-$\ell$, and
so seem disfavoured by recent observational hints for a lack of CMB power at $\ell\lesssim 40$.
We also comment on the importance of initial conditions and the possibility to have multiple pre-inflationary stages.}

\begin{document}
\hfill {DESY-14-117}

\maketitle

\section{Introduction}
\label{sec:intro}

It has been fifty years since Penzias and Wilson discovered the cosmic microwave background (CMB) radiation \cite{PW}. As a collection of free streaming photons liberated from the primordial plasma, the CMB stands as both a temporal and spatial limit to electromagnetic observations. Therefore, observing the Universe at redshifts greater than $z\sim1090$ requires either new probes or indirect signatures of primordial physics.

In the more recent past, Planck \cite{Ade:2013zuv} and BICEP2 \cite{Ade:2014xna} collaborations have generated high resolution maps of the CMB. These maps are a photograph of the plasma at decoupling, and their resolution offers precision tests of models of the early Universe.

Inflation \cite{Guth:1980zm,Starobinsky:1980te,Linde:1981mu,Albrecht:1982wi} is a successful and well-studied paradigm for generating primordial fluctuations \cite{Mukhanov:1981xt} which seed the temperature anisotropy of the CMB, galaxy formation and the general growth of large scale structure. Models of inflation are now tightly constrained by the data \cite{Ade:2013uln}. One notable constraint is the slight scale-dependence of the power spectrum of the primordial curvature perturbation. To a good approximation, the Fourier modes of this perturbation are independent quantities. Inflation predicts that the modes are generated upon leaving the horizon, and subsequently stretch along with the rest of the Universe. In short, perturbations on the largest scales are generated at the earliest time. Consistent with the `slow-roll' deformation of a de Sitter (dS) Universe, such scale dependence amounts to the time-evolution of the inflaton field.

While the prospect of primordial dynamics is exciting, the CMB itself only probes around $7$ e-foldings of expansion. While future experiments will extend this to much smaller scales, larger scales are simply impossible to measure. There is no \textit{a priori} reason why the largest scales probed by the CMB should be the first scales to have left the horizon during inflation. This is important insofar as different inflationary scenarios predict wildly different numbers of e-foldings. Despite this observational obstruction, one may use the data to regress inflationary models to earlier times. For a substantial number of additional e-foldings, this was done with scale dependent non-Gaussianity in \cite{Bramante:2013moa}. The present work takes aim at times just before the CMB scales left the horizon. In this paper we shall parametrize the space of possible dynamics as deviations to the observable scalar power spectrum at large scales.

There are two reasons such scales are of interest. The first is theoretical: inflation is hard to maintain. A large amount of inflation seems to require some fine tuning \cite{Hawking:1987bi,Gibbons:2006pa}. General arguments from the string landscape also suggest a statistical tendency towards fewer e-foldings \cite{Freivogel:2005vv}. If this is a general property of inflationary models, the most likely model of our Universe involves the minimum number of e-foldings. This would suggest interesting dynamics would occur on scales comparable to our Hubble volume. In particular, this suggests that the lowest moments of the angular power spectrum would (in addition to statistical fluctuations) encapsulate these dynamics.

The second reason is observational: temperature fluctuations at the largest scales seen in the CMB appear less correlated. This is in contrast to the expected slight enhancement of the power spectrum associated to the observed redshifting. While some astrophysical effects may influence physics at these scales \cite{Finelli:2005zc}, the perturbations at large scales, for example, are too large to be affected by the baryonic acoustic oscillations.

This `low power at large scales' anomaly was already present in COBE data \cite{Bond:1998zw} and supports a rich literature. In the remainder of this section we shall give a brief history of the study of this anomaly. Then in Sec.~\ref{sec:method} we review the analytic derivation of the perturbations and establish notation. In Sec.~\ref{sec:pre_inf} we apply this technology to parametrize the effect of pre-inflationary physics on perturbations at large angular scales. In Sec.~\ref{sec:four} we delve into two further effects which modify the large scale part of the primordial spectrum: deviations from isotropy and the Bunch-Davies (BD) vacuum. We also consider the specifics of a pre-inflationary period of curvature domination. In Sec.~\ref{sec:multiStage} we generalize this case to a one with multiple pre-inflationary stages. Finally, we conclude with a general discussion in Sec.~\ref{sec:discuss}.

\subsection*{History of the power suppression}

A lack of correlation between large-scale temperature fluctuations in the CMB was first observed as a low quadrupole moment and a lack of angular correlation beyond $\sim 60^\circ$ in COBE data \cite{Bond:1998zw,Hinshaw:1996ut}. Given the modest signal-to-noise ratio, little attention was paid to the apparent anomaly until its presence was again detected in the first year WMAP results \cite{Spergel:2003cb}. The apparent suppression of power persisted in subsequent data releases and its statistical significance, while meager, was not diminished by the release of the Planck data \cite{Ade:2013uln}.

There are a number of ways to represent these data. Typically, the moments of the angular power spectrum are shown, wherein those for $\ell \lesssim 30$ fail to be in good agreement with the inflationary predictions. The restriction of our measurements to our unique position in the Universe imparts a theoretical uncertainty for these $C_{\ell}$, of order $\sim 1/\ell$. Such \textit{cosmic variance} alone would seem to offer an explanation for the apparent discrepancy. Despite this, the collective \textit{absence} of correlation at large scales is still unexpected.

Note that the above analysis implicitly assumes the fluctuations are statistically isotropic and Gaussian, so that the angular moments of the sky $a_{\ell m}$ represent individual measurements. The study of large angular scales has revealed further statistical anomalies, notably the apparent alignment of the quadrupole and octupole moments of the angular power spectrum, \cite{Tegmark:2003ve, deOliveiraCosta:2003pu}. This suggest a possible departure from primordial isotropy, as modeled in \cite{Gumrukcuoglu:2007bx,Pereira:2007yy,Pitrou:2008gk,Gumrukcuoglu:2008gi}.

In a series of papers which followed these results, the authors of \cite{Copi:2013cya} independently investigated the statistical significance of the large angle data. This analysis emphasizes use of the $S_{1/2}$ statistic of \cite{Spergel:2003cb} as a means to quantify the apparent suppression of power, and emphasizes that the remarkable property of the CMB data is not so much that it disagrees with the Sachs-Wolfe plateau of the $\Lambda$CDM model, but rather that there is no observable correlation between fluctuations beyond $60^{\circ}$ \cite{Copi:2008hw}. Therefore, a theoretical explanation for the apparent suppression of power at large angular scales is warranted.

As yet unknown astrophysical or systematic effects may be responsible for the apparent suppression of power at large scales \cite{Copi:2010na}. The late time integrated Sachs-Wolfe effect may also affect the lowest angular moments \cite{Finelli:2005zc}. There are also a number of possible cosmological explanations; the present work aims to study their generic properties.

The main theme of such primordial explanations for power suppression at large scales is that inflation lasted for only a minimal duration. The suppression is generated by the dynamics of the condensate at the onset of inflation. Condensate dynamics have been used to model departures from a Harrison-Zeldovich-like spectrum for some time. A notable early work was Double Inflation \cite{Silk:1986vc}, which sought to explain an apparent discrepancy between cluster and galaxy correlation functions by enhancing the power at small scales while keeping the large-scale spectrum fixed.  This problem was refined to be the existence of a characteristic scale of fluctuations in the gravitational potential. To model the creation of such a scale using condensate dynamics alone, a refinement of the double inflation scenario was presented in \cite{Polarski:1992dq}. Here a period of matter domination separated the two epochs of accelerated expansion. Despite this attempt, such a scale is now known to be sourced by baryonic acoustic oscillations just prior to decoupling. Evidently, care must be taken when modeling cosmological data with primordial effects.

With the first WMAP release, the seminal work of \cite{Contaldi:2003zv} modelled the apparent suppression of power at large scales with condensate dynamics. Here inflation emerged from a period of kinetic energy domination, as per the attractor dynamics of the gravity-scalar system. This scenario was compared to WMAP1 data but the improvement of the fit to the cosmological parameters over the standard $\Lambda$CDM model was marginal \cite{Cline:2003ve}.

As the anomaly persisted through more precise data, a number of realisations of the \cite{Contaldi:2003zv} scenario were proposed \cite{Linde:1999wv,Burgess:2002ub,BasteroGil:2003bv,Bridle:2003sa,Enqvist:2004xv,Buchel:2004df,Powell:2006yg,Jain:2008dw,Bousso:2013uia,Ramirez:2012gt}. All assume `just enough' inflation as with the eponymous \cite{Schwarz:2009sj}. Perhaps owing to the original context, many authors retained a blue-shifting mechanism such as kinetic energy domination. A notable exception was \cite{Nicholson:2007by} where a period of kinetic energy domination was contrasted with a primordial epoch of radiation domination, leading to similar scalar power spectra but different predictions for the tensor modes. Yet more mechanisms arising from stringy physics prior to inflation have been proposed \cite{Dudas:2012vv,Kamenshchik:2013msa,Lello:2013mfa,Biswas:2013dry,Liu:2013iha,Kitazawa:2014dya}.

The non-observation of non-Gaussianity in the Planck analysis put many models of inflation in tension with observations. After Planck, models with concave-down scalar potentials seemed to fit the data well. These included variations of the Starobinsky model and inflection point inflation. Having `just enough' inflation while maintaining a redshifted spectrum imposes to nontrivial constraints but can be managed \cite{Downes:2012gu,Pedro:2013pba,Cicoli:2013oba}.

Recently the BICEP2 experiment \cite{Ade:2014xna} found a B-mode polarisation signal consistent primordial tensor modes. While uncertainties with the dust polarization fraction has put significance of the signal in question \cite{Flauger:2014qra,Mortonson:2014bja}, the magnitude of the BICEP2 fit to tensor-to-scalar ratio with $r\sim\mathcal{O}(0.1)$ warrants its consideration. Tensor modes contribute indirectly but positively to the temperature power spectrum at large scales \cite{Weinberg:2008zzc}. If the BICEP2 signal survives, this then deepens the gap between large scale observations and the inflationary prediction \cite{Freivogel:2014hca,Bousso:2014jca,Hu:2014aua}. The authors of \cite{Bousso:2014jca}, for example, suggest that this deviation could reach a significance at the $3.5-4\sigma$ level. Therefore dynamical power suppression --- like a period of `just enough' inflation --- offers a clean explanation for the apparent tension between the BICEP2 and Planck analyses.

In the present work we study the general properties of the scalar power spectrum at the onset of inflation. This generically leads to a change in the primordial spectrum in the form of either an abrupt enhancement or suppression of correlation on scales larger than the horizon size at the onset of inflation. If the number of inflationary e-foldings is sufficiently short we demonstrate how these effects imprint on the largest angular scales of the CMB. We illustrate this parametrization  with an example of power suppression and compare it to the anomalously low power observed at large scales in the CMB data.

\section{Pre-inflationary phases and power spectrum}
\label{sec:method}

In this section we outline the procedure that allows one to analytically determine the primordial power spectrum of scalar and tensor perturbations as well as the simplifying assumptions regarding the background evolution. The key ingredient in the analysis that will follow is the existence of multiple and well separated stages in the history of the Universe, together with the assumption that the transition between such phases is instantaneous.
In the comoving gauge, we can identify the curvature perturbation $\mathcal{R}$ and the transverse-traceless tensor perturbation $h_{ij}$,
\be
g_{ij} = a^2\left[(1-2\mathcal{R})\delta_{ij}+ h_{ij}\right].
\ee
Our focus will be primarily on $\mathcal{R}$. Quantizing $\mathcal{R}$ is made transparent by using the canonically normalized field $u=z\mathcal{R}$, with $z\equiv a \sqrt{2 \epsilon}$. The Mukhanov-Sasaki (MS) equation gives a field equation for $u$, whose Fourier-transform is
\be
u_k''+\left(k^2-\frac{z''}{z}\right)u_k=0,
\label{eq:MS}
\ee
with $\epsilon= -\frac{\dot{H}}{H^2}=-\frac{a H'}{H^2}$.
We find it useful to write the MS equation in terms of efoldings, $N_e\equiv \ln a$, as (with $X^*\equiv \partial X/\partial N_e$)
\be
u_k^{**}+ \left(1+\frac{H^*}{H}\right)u_k^*+\left[\left(\frac{k}{a H}\right)^2-4 +\frac 14 \left(\frac{H^{**}}{H^*}-3\right)^2
-\frac 14 \left(\frac{H^*}{H}-1\right)^2+\frac{H^{**}}{2 H}-\frac{H^{***}}{2 H^*}\right]u_k=0\,.
\label{eq:MSNe}
\ee
The advantage of rewriting Eq. (\ref{eq:MS}) as (\ref{eq:MSNe}) becomes evident when one considers backgrounds that are well approximated by $a H \sim e^{\xi N_e}$. This simple parametrisation allows us to model the background dynamics by changing the constant $\xi$, which is related to the equation of state parameter $w$ of a given background as
\be
\xi=-\frac12(1+3w).
\label{eq:omega}
\ee
With this parametrisation of the background dynamics we note that
\be
H^*=(\xi-1) \:H,\qquad H^{**}=(\xi-1)^2\: H,\qquad H^{***}=(\xi-1)^3\: H\,,
\ee
which simplifies the MS equation to
\be
u_k^{**}+ \xi \: u_k^*+\left[\left(\frac{k}{a H}\right)^2-(1+\xi )\right]u_k=0\,.
\label{eq:MSSimp}
\ee
As is well known, for $\xi\neq 0$ the solution to Eq. (\ref{eq:MSSimp}) is given as a sum of Hankel functions:
\be
u_k= \frac{1}{\sqrt{\xi a H}}\left[ C^{(1)}\,H_\nu^{(1)}\left(\frac{k}{\xi a H}\right)+C^{(2)}\,H_\nu^{(2)}\left(\frac{k}{\xi a H}\right)\right]\,,
\label{eq:Hankels}
\ee
with index $\nu=\left|\frac{2+\xi}{2 \xi}\right|$. For a curvature dominated background, $\xi=0$, the solution to (\ref{eq:MSSimp}) takes the form
\be
u_k= C^{(1)}\,e^{\sqrt{1-(k/H_0)^2}N_e}+C^{(2)}\,e^{-\sqrt{1-(k/H_0)^2}N_e}\,,
\label{eq:HankelsCurvature}
\ee
where $H_0$ is the value of the Hubble parameter at the onset of curvature domination. In both cases the constants $C^{(1)}$ and $C^{(2)}$ are functions of the comoving momentum $k$, to be determined by the boundary conditions of the problem.

We are interested in backgrounds that undergo multiple phases with distinct dynamics: an example of this would be a Universe in which two inflationary phases ($w=-1\,\Leftrightarrow\,\xi=1$) are separated by a matter dominated period ($w=0\,\Leftrightarrow\,\xi=-1/2$) \cite{Polarski:1992dq} or perhaps one undergoing a period of fast-roll ($w=1\,\Leftrightarrow\,\xi=-2$) followed by a slow-roll phase \cite{Contaldi:2003zv}.

At this stage we do not assume any particular sequence of events in the pre-inflationary history of the Universe and will generically take the Hubble parameter to be a continuous function defined as
\be
H=\sum_i H_i \:e^{(\xi_i-1)N_e} \:\Theta(N_{e,i},N_{e,i+1}),
\ee
where $\Theta(N_{i},N_{i+1})$ is the unit window function of the interval $[N_{e,i},N_{e,i+1}]$ and $H_i$ is the value of the Hubble parameter at $N_e=N_{e,i}$. The $H_i$ are determined by requiring the continuity of $H$ across the various transitions and can be expressed recursively as
\be
H_{i+1}=H_i \: e^{(\xi_i-\xi_{i+1})N_{e,i}}.
\ee

\begin{figure}[t!]
	\centering
	\begin{minipage}[b]{0.49\linewidth}
	\centering
	\includegraphics[width=1\textwidth]{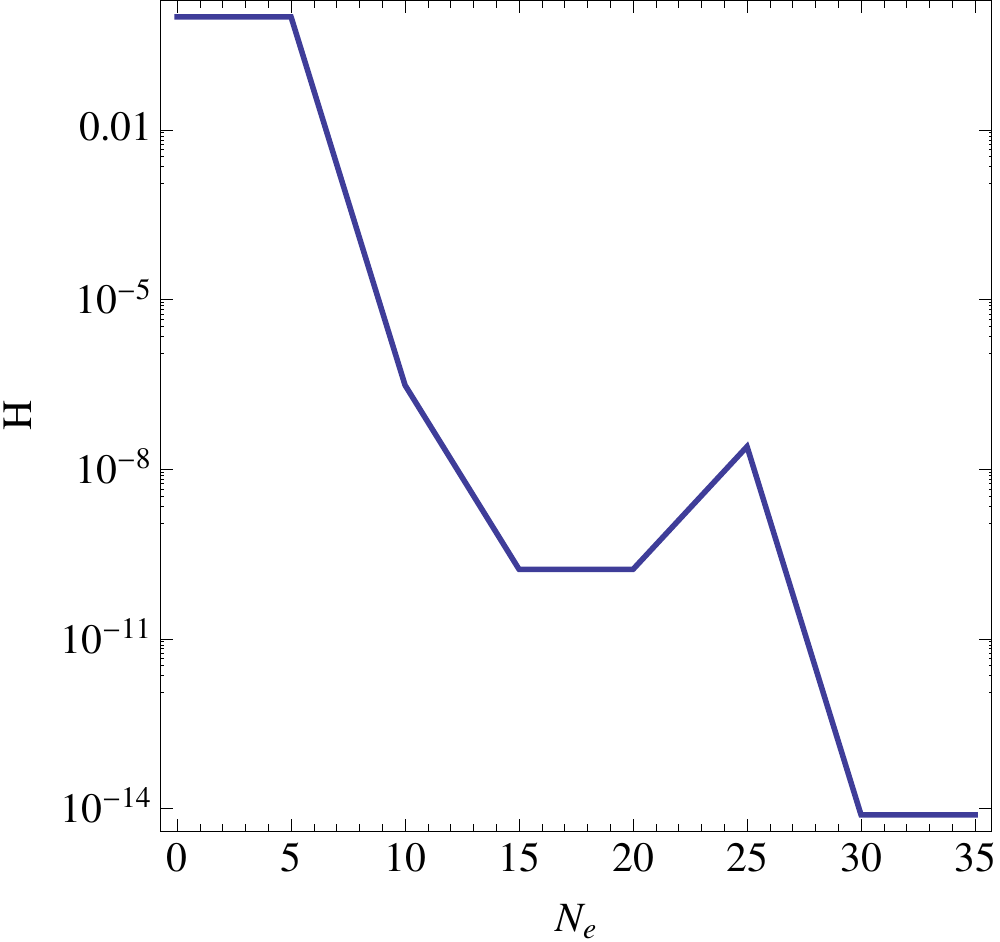}
    \end{minipage}
	\hspace{0.05cm}
	\begin{minipage}[b]{0.49\linewidth}
	\centering
\includegraphics[width=0.95\textwidth]{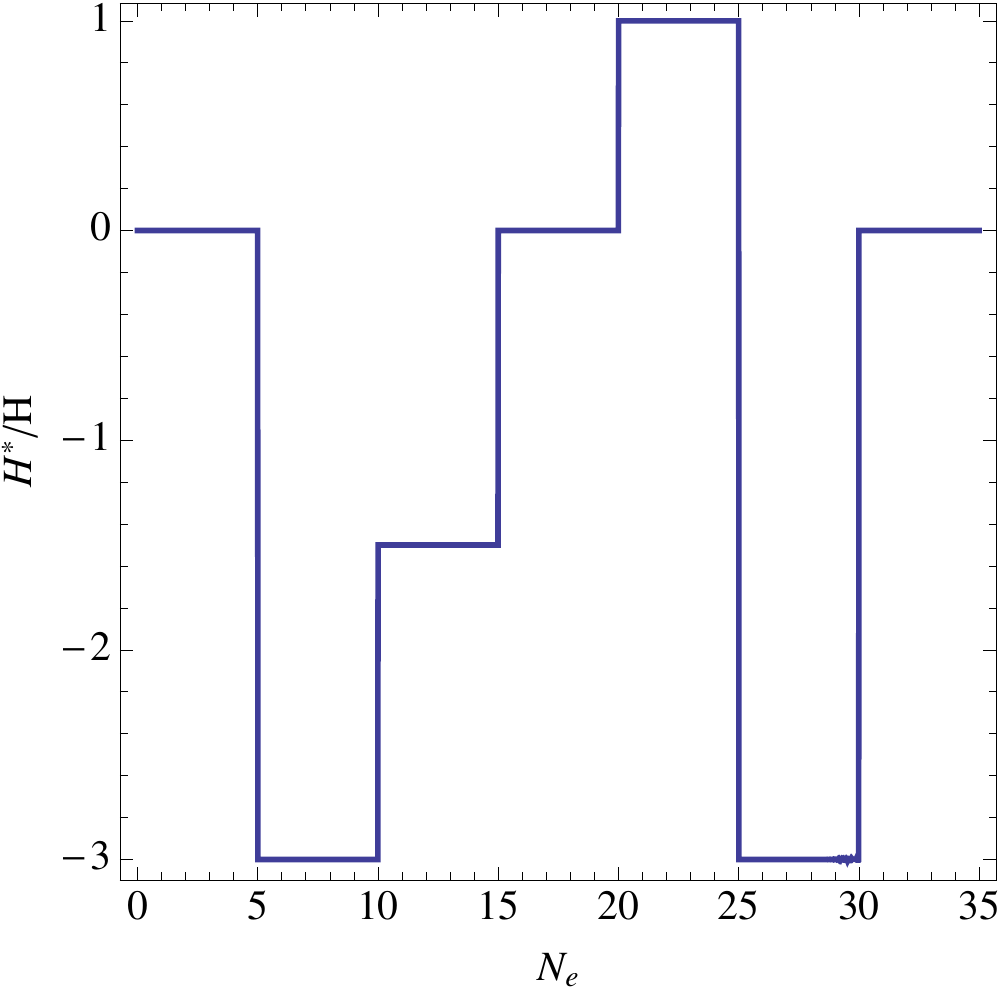}
	\end{minipage}
	\hspace{0.05cm}
\caption{$H$ (left) and  $H^*/H$ (right) for a Universe undergoing the unlikely sequence of events: inflation $\rightarrow$ fast-roll $\rightarrow$ matter domination $\rightarrow$ inflation $\rightarrow$ super-inflation $\rightarrow$ fast-roll $\rightarrow$ inflation.}
	\label{fig:DHHtoy}
\end{figure}

Solving the MS equation in such a background is similar to solving the Schr\"odinger equation across a succession of potential barriers and potential wells with different heights and depths. This analogy becomes apparent when one looks at the behaviour of $H^*/H$ (which determines the behaviour of $z''/ z$) for a Universe undergoing multiple transitions which we illustrate in Fig. \ref{fig:DHHtoy}.

In each stage of the history of the Universe, the curvature perturbations are determined by Eqs. (\ref{eq:Hankels}) and (\ref{eq:HankelsCurvature}). Similar to quantum mechanical problems, the integration constants $C_i^{(1)}, C_i^{(2)}$ of the $i$-th phase are determined by requiring continuity of the `wave function' $u_k$ and of its derivative across the various transitions. In this way one can analytically determine the shape of the primordial power spectrum\footnote{We choose to compute $P_Q$ instead of $P_{\zeta}$ for several reasons. Firstly, because $P_T\sim P_Q$ with an ${\cal O}(1)$ proportionality factor, we compute the tensor power spectrum as well while determining $P_Q$. Next, for equations of state where the curvature perturbation $\zeta$ exists and no strong changes of the slow-roll parameter $\epsilon$ occur, $P_\zeta$ and $P_T$ are qualitatively similar since they just come with a slow-varying ratio controlled by $\epsilon$. Hence, in these cases computing $P_Q$ gives us the relevant information about both $P_\zeta$ and $P_T$. Finally, there are cases like false-vacuum dS space, where $\zeta$ does not exist. Consequently, there is no $P_\zeta$ but there are still tensor perturbations, and consequently computing $P_Q$ still gives us $P_T$. Moreover, tensor modes do convert at $\ell < 50$ into a contribution $(\delta C_\ell^{TT})_{\rm tensor}$ to the temperature two-point function $C_\ell^{TT}$ power spectrum. Hence, computing $P_Q$ gives us information about the asymptotic behaviour of the temperature $C_\ell$'s at low-$\ell$ even in cases where $P_\zeta$ does not exist.}
\be
P_Q=k^3 \left|\frac{u_k}{a}\right|^2\,,
\ee
once the initial conditions and the background solution are specified. In Appendix \ref{App} we present the recursive relations that allow for the  analytical determination of the primordial spectrum.

For concreteness we focus on $\xi\neq0$ backgrounds in what follows, keeping in mind that including $\xi=0$ periods requires only minor modifications of the method to be described. The primordial power spectrum $P_Q$ is to be evaluated at horizon crossing $k=aH$ for each mode $k$. We can use the asymptotic expansion of the Hankel functions
\be
\lim_{x\rightarrow0} H_\nu^{(1,2)}(x)=\mp i \frac{2^{-\nu } \cos\left(\pi  \nu \right) \Gamma(-\nu )}{\pi }\ x^{\nu } \mp i \frac{ 2^{\nu }  \Gamma(\nu )}{\pi }\ x^{-\nu }+\frac{2^{-\nu }}{\Gamma(1+\nu )}\ x^{\nu }
\ee
to write the late time spectrum as
\begin{eqnarray}
P_Q&=&\left(\frac{k}{a H}\right)^3 H^2 \left\{ \left(\frac{k}{a H}\right)^{-2 \nu }\left|C^{(1)}-C^{(2)}\right|^2
\frac{4^{\nu }  |\xi|^{-1+2 \nu }\Gamma(\nu )^2}{\pi ^2}\right. \nonumber \\
&+&\left(\frac{k}{a H}\right)^{2 \nu }\left|-i \left(C^{(1)}+C^{(2)}\right)+\left(C^{(1)}-C^{(2)}\right) \cot(\pi\nu)\right|^2
\frac{4^{-\nu }   |\xi|^{-1-2 \nu } }{\Gamma(1+\nu)^2} \nonumber \\
&+& \left.\left[i \left(\overline{C}^{(1)}C^{(2)}-C^{(1)} \overline{C}^{(2)}\right)-\left|C^{(1)}-C^{(2)}\right|^2 \cot(\pi \nu)\right]
\frac{2}{\pi |\xi|\nu }\right\}\,,
\label{eq:PQ}
\end{eqnarray}
where $\xi$ and $\nu$ refer to the last phase, $i=i_{\rm max}$ and consequently $C^{(1)}=C^{(1)}_{i_{\rm max}}$ and $C^{(2)}=C^{(2)}_{i_{\rm max}}$.
Driven by observational evidence, we will be interested in computing the spectrum after a phase of inflation ($\nu_{i_{\rm max}}=3/2$) with the approximate duration of 60 efoldings, while allowing for the existence of putative pre-inflationary phases. From Eq. (\ref{eq:PQ}) it follows that the effect of any pre-inflationary phases is fully encoded in the non-trivial nature of the integration constants $C_{i_{\rm max}}^{(1)}$ and $C_{i_{\rm max}}^{(2)}$.

\begin{figure}[h!]
 \centering
	\includegraphics[width=0.5\textwidth]{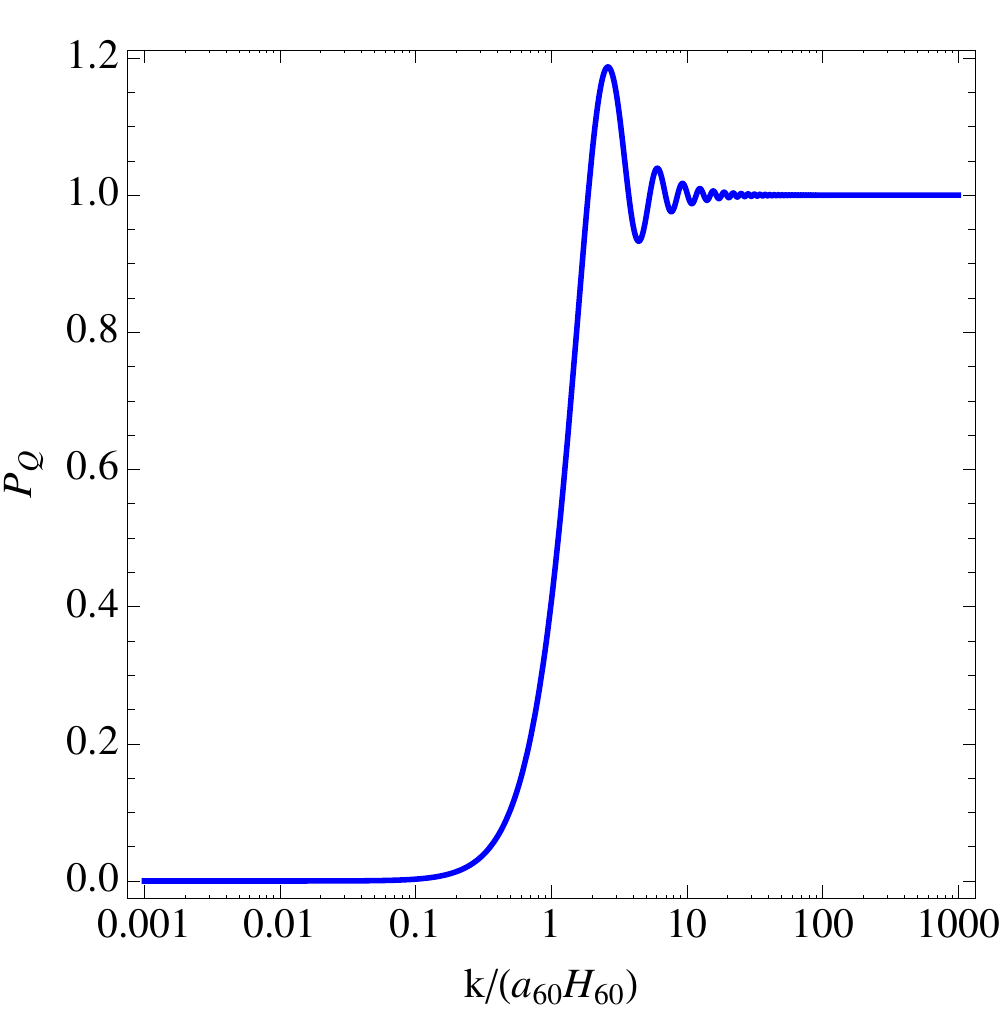}
 \caption{Power spectrum for a Universe undergoing a fast-roll followed by a dS phase. Note that the suppression of power happens on scales that were outside the horizon for the whole duration.}\label{fig:Fast}
\end{figure}

As an example, let us consider a Universe that underwent a period of fast-roll followed by a slow-roll phase \cite{Contaldi:2003zv}.
Moreover let us assume that the initial conditions for the density perturbations in the pre-inflationary phase asymptote to the BD vacuum deep inside the horizon which selects one of the two Hankel functions: $C^{(1)}_1=\sqrt{\pi/2}$ and $C^{(2)}_1=0$. In this case, direct application of the results in Appendix \ref{App} leads to (with $r\equiv k/(a_{60}H_{60})$)
\begin{eqnarray}
C^{(1)}_2 &=&  -\frac{1}{8} \pi ^{3/2} r \left[H_0^{(1)}\left(-\frac{r}{2}\right)
H_{\frac{1}{2}}^{(2)}(r)+H_1^{(1)}\left(-\frac{r}{2}\right)
H_{\frac{3}{2}}^{(2)}(r)\right] \\
C^{(2)}_2 &=& \frac{1}{8} \pi ^{3/2} r \left[H_0^{(1)}\left(-\frac{r}{2}\right)
H_{\frac{1}{2}}^{(1)}(r)+H_1^{(1)}\left(-\frac{r}{2}\right)
H_{\frac{3}{2}}^{(1)}(r)\right]\,.
\end{eqnarray}
The power spectrum is then modulated by $\left|C^{(1)}_2-C^{(2)}_2\right|^2$ which displays the following limiting behaviour
\be
\left|C^{(1)}_2-C^{(2)}_2\right|^2\rightarrow \left\{
\begin{array}{c}
0 \qquad\text{for}\qquad k/ (a_{60}H_{60})\rightarrow 0\\
1 \qquad\text{for}\qquad k/ (a_{60}H_{60})\rightarrow \infty\\
\end{array}\right.
\ee
and so we see that there is a suppression of power for modes that were superhorizon at the onset of the dS stage $k\ll a_{60}H_{60}$, while on smaller scales ($k\gg a_{60}H_{60} $) the spectrum tends to the scale invariant one, characteristic of dS space. Fig. \ref{fig:Fast} depicts the behaviour of the primordial power spectrum in such a background.

With the method in place, we are now in a position to discuss in a systematic manner the implications of having `just enough' inflation for the primordial power spectrum and how the pre-inflationary history can affect it on the largest observable scales.

\section{Systematic analysis of low-$\ell$ power spectrum}
\label{sec:pre_inf}

If we take the hints from WMAP and Planck data concerning a lack of TT power at $\ell < 40$ seriously, the question of a general mechanism for such deviations from near scale-invariance arises. In this section we will show that a fairly general analysis of the effects at low-$\ell$ uses just two ingredients:
\begin{itemize}
\item The visibility of any deviation from scale invariance at large scales requires `just enough' inflation, meaning $N_e\lesssim 50 - 60$ since this number of e-foldings corresponds to the largest observable scales in the CMB sky. Hence, immediately prior to our observable inflationary phase there was a non-slow-roll phase of `pre-inflation'.

\item Any effect at low-$\ell$ arises from fluctuation modes of the order of or larger than the horizon scale at the onset of our final observable phase of slow-roll inflation.
\end{itemize}

\begin{figure}[h!]
 \centering
	\includegraphics[width=0.8\textwidth]{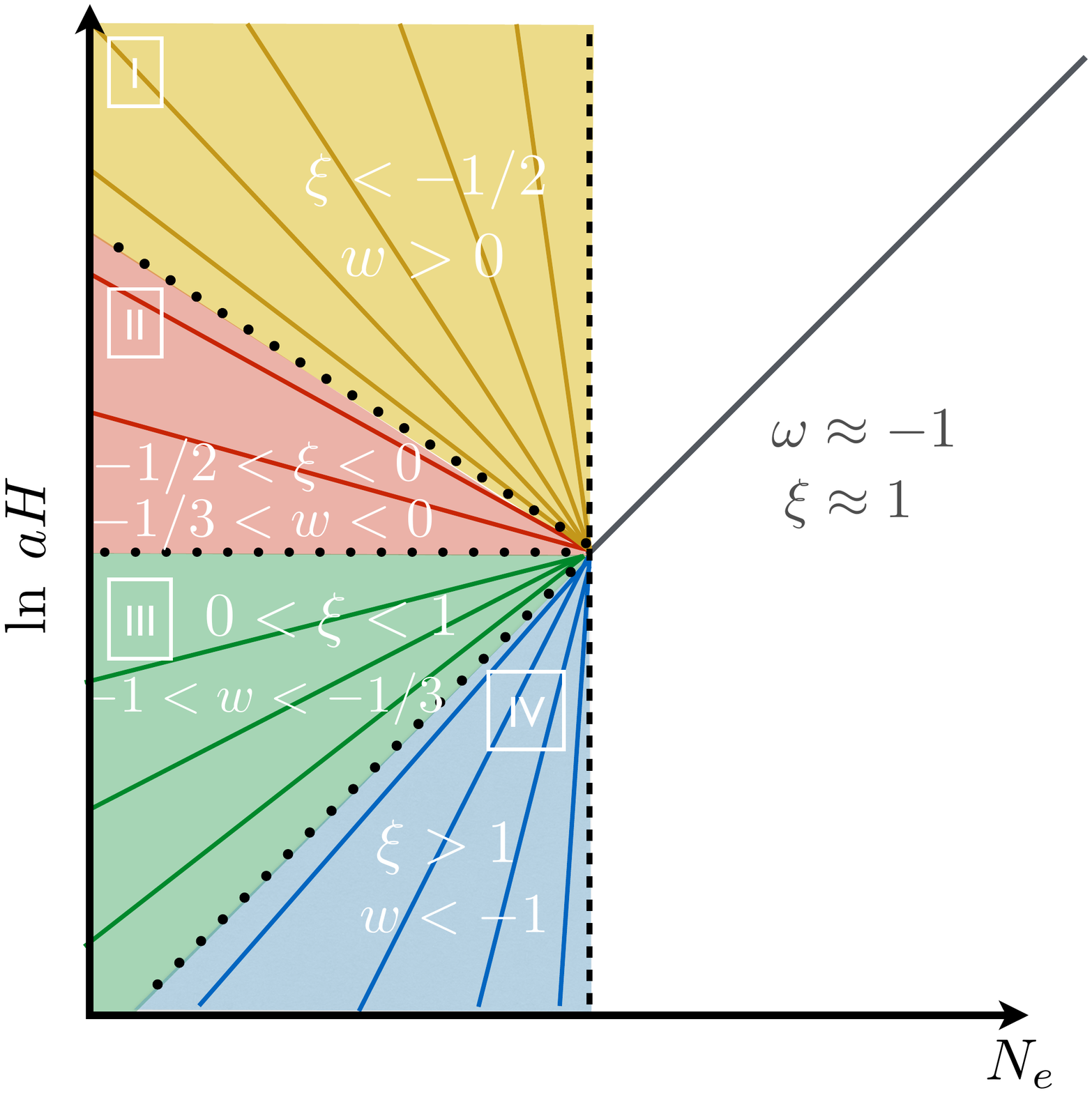}
 \caption{Pre-inflationary horizon in a logarithmic plot of the comoving horizon $\ln aH$ as a function of time measured in terms of e-foldings $N_e=\int_0^tH\,\d t$.}\label{fig:accDec}
\end{figure}

We will argue that there is a level of universality in the large scale power spectrum, in the sense that all possible background dynamics in the pre-inflationary era lead to only two distinct forms of the primordial spectrum. To make this behaviour apparent it is convenient to classify the pre-inflationary backgrounds based on their $\xi$ parameter, or equivalently on the equation of state parameter $w$, related to $\xi$ via Eq. (\ref{eq:omega}). We therefore define four regions in the $\xi$ parameter space:
\begin{itemize}
\item Region I ($\xi<-1/2 \Leftrightarrow w>0$): the horizon size is increasing and the large scale part of the primordial power spectrum is generated by modes that were superhorizon during this pre-inflationary phase. Background dynamics which fall into this region involve fast-roll ($\xi=-2\Leftrightarrow w=1$) and radiation dominance ($\xi=-1\Leftrightarrow w=1/3$). The case of matter dominance ($\xi=-1/2\Leftrightarrow w=0$) is at the boundary between regions I and II.
\item Region II ($-1/2<\xi<0\Leftrightarrow -1/3<w<0$): the horizon is increasing and the largest observable scales originate from modes that were always on superhorizon scales. The case of curvature dominance ($\xi=0\Leftrightarrow w=-1/3$) is at the boundary between regions II and III.
\item Region III ($0<\xi<1 \Leftrightarrow -1<w<-1/3$): the horizon size is decreasing albeit at a slower rate than in slow-roll inflation. The small $k$ modes of the spectrum leave the horizon during this phase.
\item Region IV ($\xi>1 \Leftrightarrow w<-1$): in this case the pre-inflationary horizon is decreasing and it is doing so at a faster rate than in the
slow-roll phase. Modes in the low-$\ell$ region left the horizon during such a phase. A typical example of background dynamics of this type is super-inflation ($\xi=2\Leftrightarrow w=-5/3$).
\end{itemize}
Fig. \ref{fig:accDec} depicts the evolution of the inverse horizon size $\left(a H\right)^{-1}$ during inflation and its preceding phase.
In what follows we will determine the primordial power spectrum for the four cases and witness the emergence of clear similarities between regions I and IV, and regions II and III, implying a high level of degeneracy in the power spectrum. To build an intuitive picture of the process, we will follow the evolution of superhorizon perturbations both during pre-inflation and inflation, using the results of Sec. \ref{sec:method}, in particular Eq. (\ref{eq:PQ}).

\subsection{Pre-inflationary region I}

\begin{figure}[t!]
 \centering
	\includegraphics[width=0.80\textwidth]{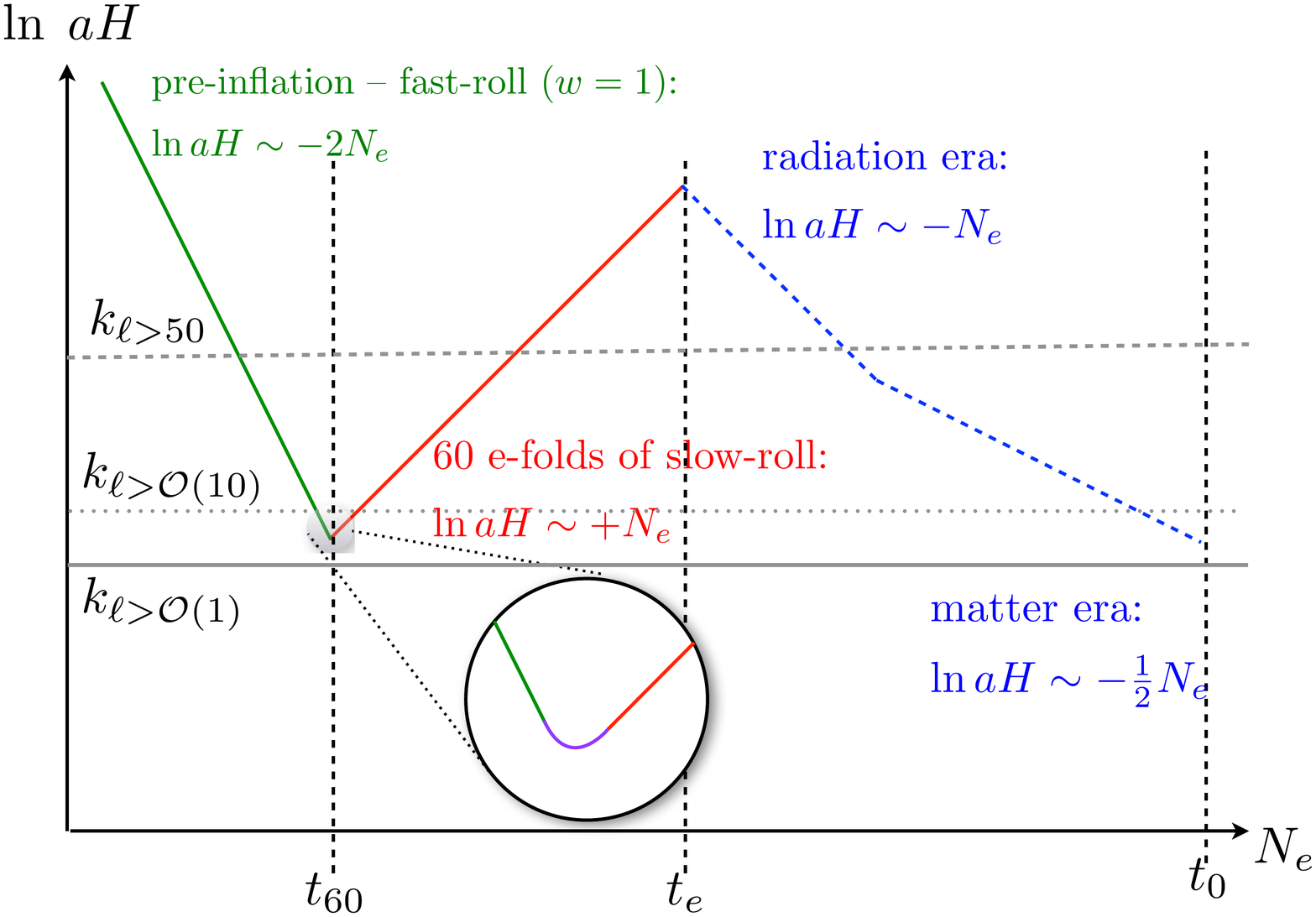}
 \caption{Logarithmic plot of the comoving horizon $\ln aH$ as a function of $N_e$. $t_0$ denotes today, $t_e$ the end of the final observable phase of slow-roll inflation, and $t_{60}$ the time corresponding to about 60 e-foldings before the end of the final slow-roll phase, corresponding to the largest observable scales in the CMB today. `Just enough' inflation means that $t_{60}$ bounds the observable slow-roll phase to the past, i.e. scales which left the horizon just around $t_{60}$ are re-entering our horizon just now around $t_0$. There is a single stage of non-slow-roll pre-inflation to the past of $t_{60}$, which we exemplify here by a fast-roll phase: $w= 1\,\Leftrightarrow\,\xi=-2$. The gray horizontal lines denote fluctuation modes with three different values of the comoving wavenumber $k$. The mode $k_{\ell>{\cal O}(1)}$ never enters the horizon during the pre-inflation or slow-roll stage. Hence, it bears memory of the quantum fluctuation of the earliest phase, the pre-inflationary one, all the way into the CMB.}\label{fig:1}
\end{figure}

We start the discussion of the primordial power spectrum by considering the evolution of the comoving horizon for the simplest situation of a single non-slow-roll stage of pre-inflation as depicted in Fig.~\ref{fig:1}. The comoving horizon evolves from a single type I pre-inflationary stage through the approximately $60$ e-foldings of `just enough' slow-roll inflation into the final radiation and matter dominated stages.
As mentioned in Sec. \ref{sec:method}, we approximate each stage by a spatially flat FRW Universe filled by a dominant perfect fluid component with a given constant equation of state $w$. The comoving horizon then obeys during each stage
\be\label{eq:c-hor}
\ln aH=\text{const}-\frac12(1+3w) N_e.
\ee
We match $\ln aH$ at each boundary between adjacent stages continuously and instantaneously as mentioned in Sec. \ref{sec:method}. This approximates the situation of a rapid turnover phase interpolating quickly between the asymptotic behaviour of two given adjacent stages (the real if rapid turnover phase being smooth as seen in the `lens' in Fig.~\ref{fig:1}).

The approximation of a rapid and hence almost instantaneous and continuous transition tells us that the quantum fluctuations present during the early stage are projected onto the new mode expansion basis of the subsequent stage immediately after the transition. This `sudden approximation' of quantum mechanics is captured by the Bogoliubov coefficients relating pre-inflationary and inflationary quantum vacua, and consequently works best with almost instantaneous transitions. The slower and smoother the actual transition happens, the smaller the effects of the changing mode expansion basis are. Hence, the use of the `sudden approximation' to calculate the Bogoliubov coefficients provides with an upper bound on the strength of ensuing effects.

For the simple two-stage history under discussion -- decelerating pre-inflation transitioning suddenly into slow-roll -- we can see the qualitative features of the fluctuation mode evolution by plotting fluctuations with different comoving wave-numbers $k$ together with the comoving horizon, as in Fig.~\ref{fig:1}.  The most relevant scale for this problem is set by the horizon size at the onset of inflation: $a_{60} H_{60}$. We see that modes corresponding to $C_\ell$'s with $\ell>{\cal O}(1)$ in the CMB are modes which have $k\lesssim a_{60}H_{60}$. By this condition these modes are super-horizon during the whole time of pre-inflation and subsequent slow-roll, re-entering the horizon only at late times, long after reheating.

We can understand the evolution of such modes in the pre-inflationary region I, during which they are outside the horizon, by rewriting Eq. (\ref{eq:PQ}) as
$P_Q\sim P_Q^1+P_Q^2+P_Q^3$ with
\begin{eqnarray}
P_Q^1 &=& \left(\frac{k}{a H}\right)^{3-2\nu} H^2\left|C^{(1)}-C^{(2)}\right|^2 \label{PQ1} \\
P_Q^2 &=& \left(\frac{k}{a H}\right)^{3+2\nu} H^2\left|-i \left(C^{(1)}+C^{(2)}\right)+\left(C^{(1)}-C^{(2)}\right) \cot(\pi  \nu )\right|^2 \label{PQ2} \\
P_Q^3 &=& \left(\frac{k}{a H}\right)^3 H^2\left[i \left(\overline{C}^{(1)}C^{(2)}-C^{(1)} \overline{C}^{(2)}\right)-\left|C^{(1)}-C^{(2)}\right|^2 \cot(\pi \nu)\right]\,, \label{PQ3}
\end{eqnarray}
where we have neglected constant prefactors in each term.
Depending on $\xi$, the different contributions to $P_Q$ display distinct behaviours. For $\xi<-2$ one finds $P_Q^1\propto \left(a H\right)^0$
while $P_Q^2 ,P_Q^3 \propto \left(a H\right)^{-1}$. Recalling that for this family of background solutions $a H$ is decreasing exponentially with $N_e$,
we realise that $P_Q^1$ corresponds to a frozen mode whereas $P_Q^2+P_Q^3$ give rise to a growing mode. For $-2<\xi<0$ backgrounds
the same reasoning implies that $P_Q^1+P_Q^3$ form a decaying mode while $P_Q^2$ is frozen. Provided there are no special cancellations occurring in the $C^{(1)}$ and $C^{(2)}$ dependent prefactors\footnote{An interesting exception is the case of radiation followed by matter domination after an initial phase of slow-roll inflation. In this case the continuity of the perturbations across both transitions implies, to a very good approximation, $C^{(1)}=C^{(2)}$ both in the radiation and matter eras. This  has the effect of killing the dominant decaying solution leaving behind a spectrum of  frozen perturbations on superhorizon scales.} we expect the constant solution to dominate for $\xi<-2$, since
\be
\frac{P_Q^1}{P_Q^2+P_Q^3}\sim \left(\frac{k}{ a  H}\right)^{-2\nu}\ll 1\,,
\ee
and the decaying to dominate for $-2<\xi<0$ backgrounds given that
\be
\frac{P_Q^1+P_Q^3}{P_Q^2}\sim \left(\frac{k}{ a  H}\right)^{-2\nu}\ll1\,.
\ee
The spectral index of the dominant solution, provided by $P_Q^1$ in both regimes, can be read off from Eq. (\ref{PQ1})
\be
\left.n_s-1\right|_{\rm pre-inf}\equiv 3-2 \nu_{\rm pre-inf} >0\,,
\ee
and so the spectrum of $k<a_{60} H_{60}$ modes during the pre-inflationary epoch is blue tilted: there is less power on the largest length scales. When these modes cross into the slow-roll phase they will freeze (if they were not already frozen) and this blue tilt for low-$k$ modes will remain a feature of  the late time power spectrum.

The $k_{\ell>{\cal O}(10)}$ modes shortly re-enter the horizon close to the end of pre-inflation and leave again shortly after the beginning of slow-roll. Consequently, these modes undergo some evolution while inside the horizon which will show up as interference via oscillating Bogoliubov coefficients.

Finally, the $k_{\ell>50}$ modes stay deep inside the horizon for some time before leaving again during slow-roll. Therefore, even though they lived through the pre-inflationary phase,  these modes evolve like Minkowski QFT fluctuations while deep inside the horizon and forget the preceding non-slow-roll phase. When they leave the horizon during slow-roll, they just form the BD vacuum for a quasi-dS stage and hence asymptote to the known quasi scale-invariant power spectrum in the UV: $P_Q\sim H^2\ k^{n_s-1}$ with $n_s=1+\mathcal{O}(\epsilon,\eta)$.

We now have all the information to determine how the IR and UV power spectrum compare at the end of inflation. For that one may take the ratio
\be
\frac{P_Q^{IR}}{P_Q^{UV}}\equiv\frac{P_Q^{k\ll a_{60} H_{60}}}{P_Q^{k\gg a_{60} H_{60}}}\sim \left(\frac{k_{IR}}{a H}\right)^{3-2 \nu_{\rm pre-inf}} \left(\frac{H_{\rm pre-inf}}{H_{\rm inf}}\right)^2 \left(\frac{a H}{k_{UV}}\right)^{3-2 \nu_{\rm inf}}.
\ee
Noting that the quasi scale invariant nature of slow-roll inflation implies that the last factor is close to unity and that  for type I backgrounds $H_{\rm pre-inf}\ge H_{\rm inf}$, we see that the pre-inflationary spectral index ($\nu_{\rm pre-inf}<3/2$) uniquely predicts
\be
\lim_{k_{IR}\rightarrow 0}\ \frac{P_Q^{IR}}{P_Q^{UV}}\ll 1\,,
\ee
showing that in the deep IR ($k\ll a_{60} H_{60}$) there is always suppression of power relative to the UV ($k\gg a_{60} H_{60}$). We stress that the key quantity that leads to this result is the pre-inflationary spectral index which for these backgrounds is related to the choice of initial conditions for the curvature perturbations.

\begin{figure}[h!]
	\centering
	\begin{minipage}[b]{0.4\linewidth}
	\centering
\includegraphics[width=1\textwidth]{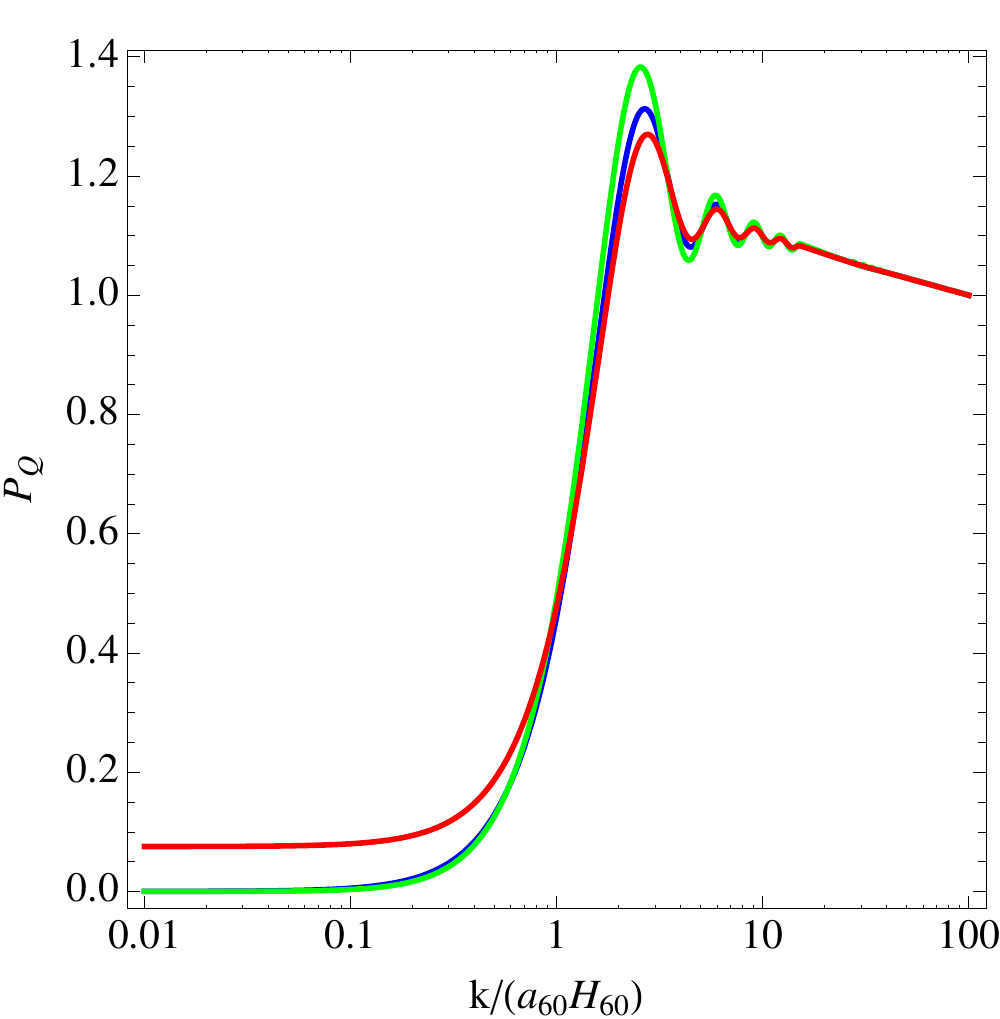}
    \end{minipage}
	\hspace{0.5cm}
	\begin{minipage}[b]{0.415\linewidth}
	\centering
\includegraphics[width=1\textwidth]{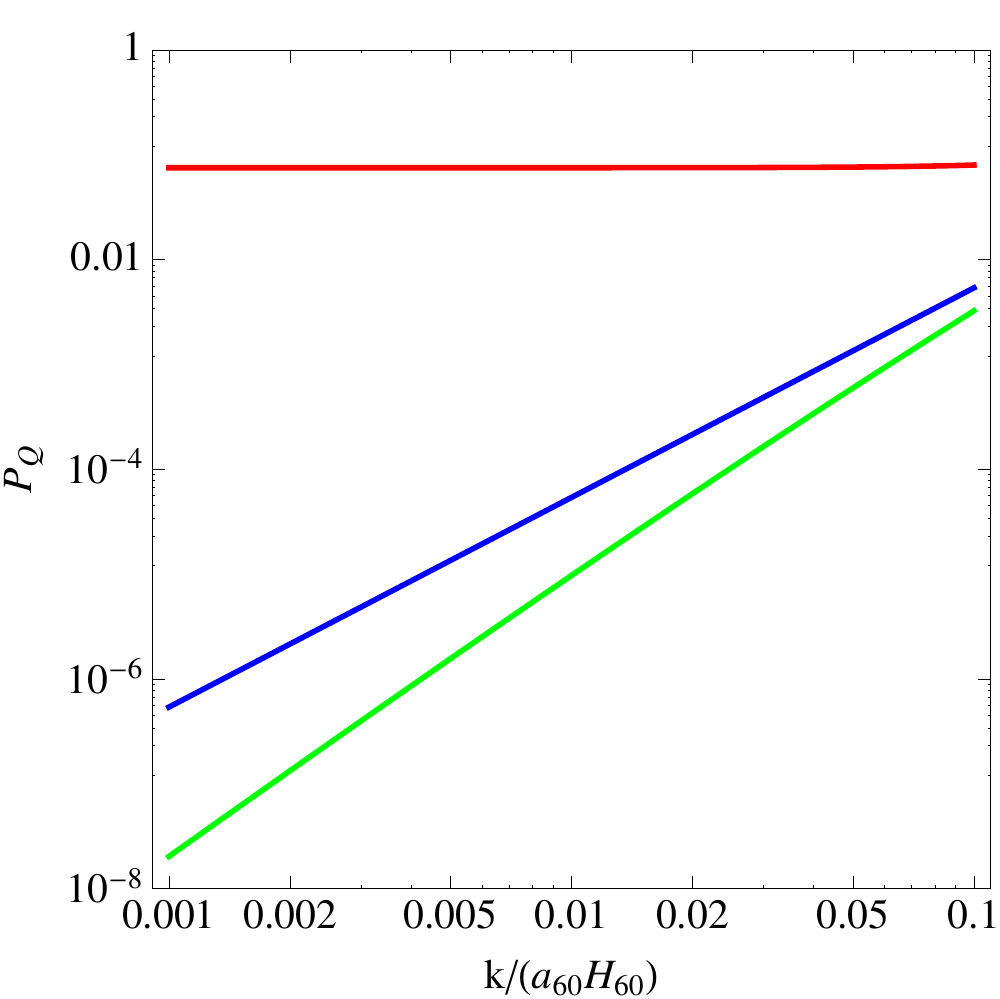}
	\end{minipage}
	\caption{Power spectrum for different type I backgrounds: matter dominance (red), fast-roll (green) and radiation dominance (blue).
The right panel displays the low-$k$ behaviour of the different solutions, their $k$ scaling being given by $k^{3-2\nu_{\rm pre-inf}}$.}
	\label{fig:type_I}
\end{figure}

These analytical results can be confirmed by an explicit computation of the primordial power spectrum for different pre-inflationary setups using the matching method of Sec. \ref{sec:method}. As an example we consider three different type I backgrounds: radiation dominance ($\xi=-1\Leftrightarrow w=1/3$ and $\nu_{\rm pre-inf}=1/2$), a fast-rolling scalar field ($\xi=-2\Leftrightarrow w=1$ and $\nu_{\rm pre-inf}=0$) and matter dominance ($\xi=-1/2\Leftrightarrow w=0$ and $\nu_{\rm pre-inf}= 3/2$). The spectra are plotted in Fig. \ref{fig:type_I}. The similarities between them are evident: they all peaking at scales near the Hubble at the onset of inflation $H_{60}$, displaying lack of power in the IR, they all have the same oscillatory frequency around $k=a_{60} H_{60}$ and the same asymptotic UV behaviour. Even though we are comparing the two-point function in only three different backgrounds, any type I background generates similar spectra, where the power loss region coincides with the modes that were outside the horizon both during the pre-inflationary and the slow-roll epoch.

\subsection{Pre-inflationary region II}

\begin{figure}[h!]
	\centering
	\begin{minipage}[b]{0.415\linewidth}
	\centering
\includegraphics[width=1\textwidth]{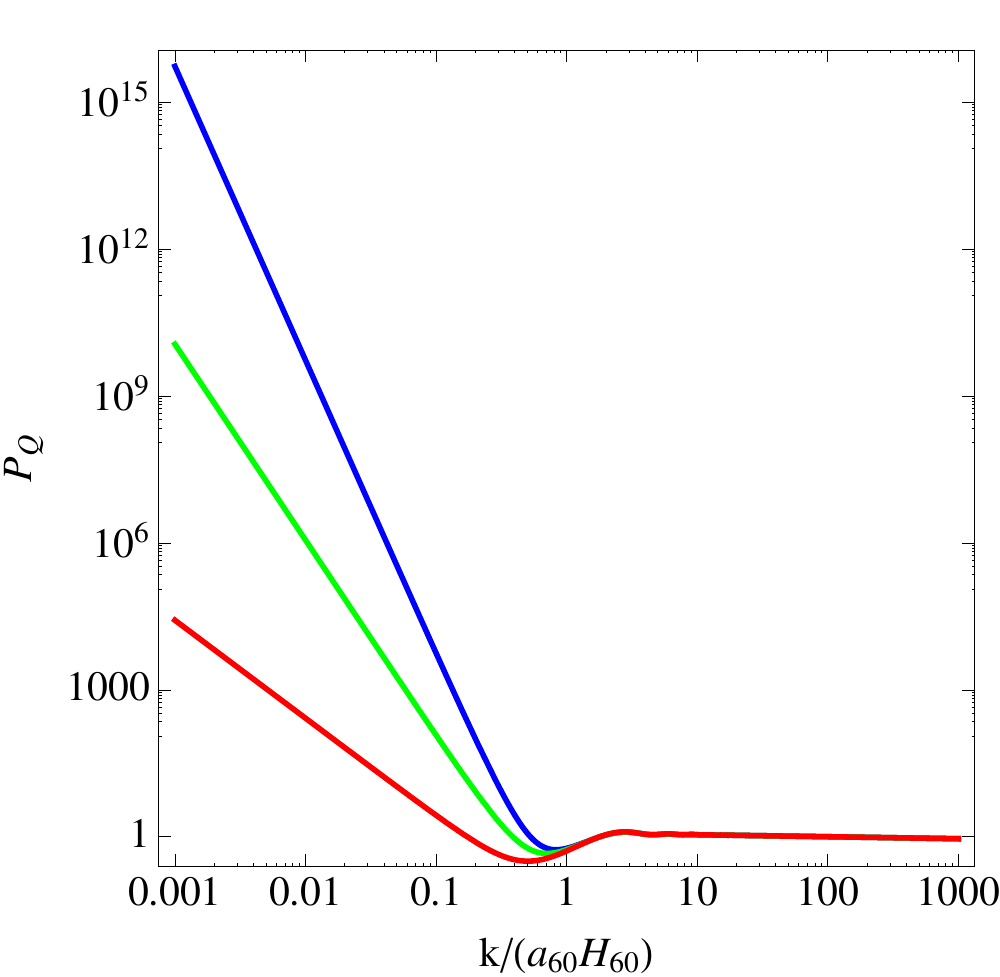}
    \end{minipage}
	\hspace{0.5cm}
	\begin{minipage}[b]{0.4
\linewidth}
	\centering
\includegraphics[width=1\textwidth]{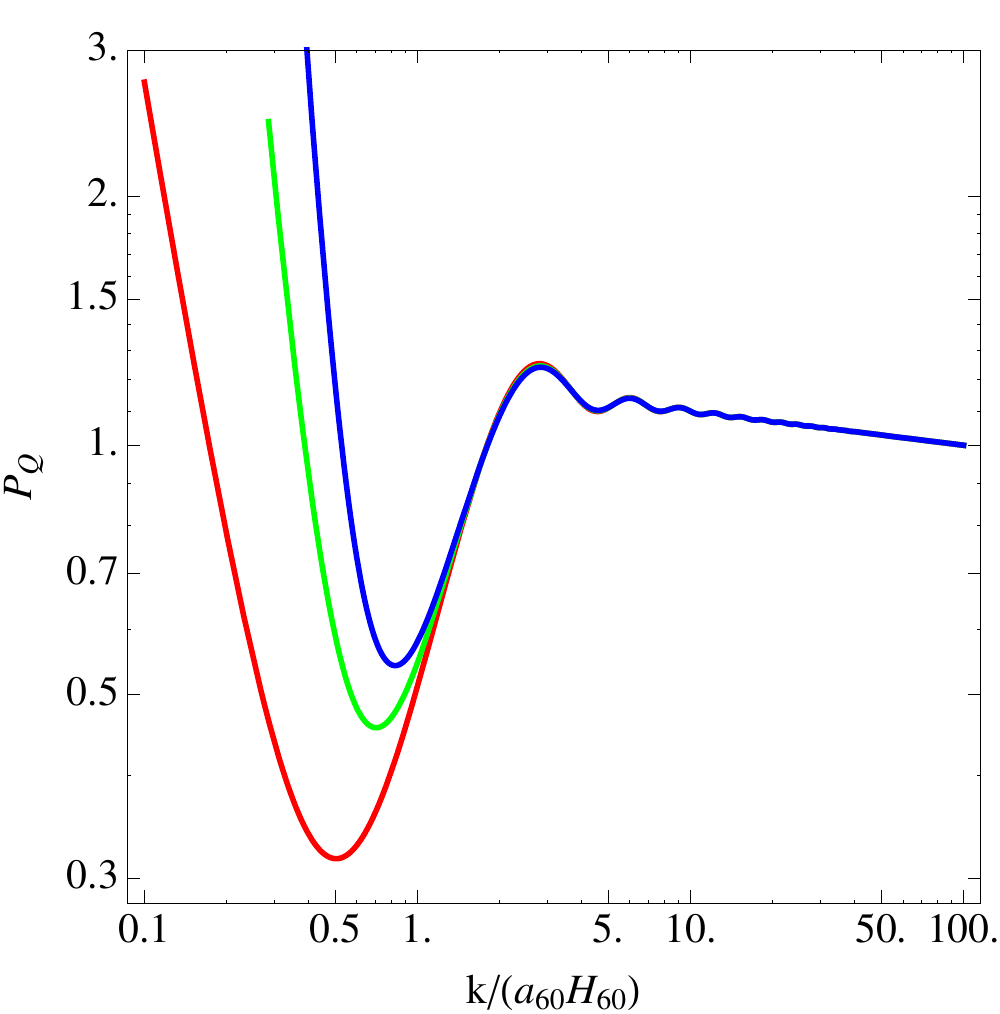}
	\end{minipage}
	\caption{Power spectrum for different type II backgrounds: $\xi=-1/3$ (red), $\xi=-1/4$ (green) and $\xi=-1/5$ (blue).
The power on large scales is enhanced due to a red spectral index on those scales.}
	\label{fig:type_II}
\end{figure}

Backgrounds of type II undergo decelerated expansion and the curvature perturbation modes involved in the IR part of the spectrum were
always outside the horizon as in the type I case. To understand the evolution of such modes during pre-inflation we can use Eqs. (\ref{PQ1})-(\ref{PQ3})
to find that for these backgrounds the spectrum is composed of a decaying and a frozen solution, given by $P_{Q}^1+P_Q^3$ and $P_Q^2$ respectively. If there are no cancellations in the $C^{(1)}$ and $C^{(2)}$ dependent factors, the decaying solution dominates and the spectrum for modes $k\ll a_{60} H_{60}$ becomes
\be
P_Q\sim H_{\rm pre-inf}^2\  \left(\frac{k}{a H}\right)^{\left.n_s-1\right|_{\rm pre-inf}}
\ee
where
\be
\left. n_s-1\right|_{\rm pre-inf}\equiv 3-2 \nu_{\rm pre-inf}<0\,.
\ee
As these modes transition to the inflationary phase they will remain frozen until they re-enter the horizon, during the post-inflationary evolution.
The modes with large enough comoving momentum $k\gg a_{60} H_{60}$ have a spectrum characteristic of slow-roll inflation, and so we find that the ratio of IR to UV power is
\be
\frac{P_Q^{IR}}{P_Q^{UV}}\equiv\frac{P_Q^{k\ll a_{60} H_{60}}}{P_Q^{k\gg a_{60} H_{60}}}\sim \left(\frac{k_{IR}}{a H}\right)^{3-2 \nu_{\rm pre-inf}} \left(\frac{H_{\rm pre-inf}}{H_{\rm inf}}\right)^2 \left(\frac{a H}{k_{UV}}\right)^{3-2 \nu_{\rm inf}}.
\ee
Given that $(k_{UV}/aH)^{3-2\nu_{\rm inf}} \sim 1$ and $\nu_{\rm pre-inf}>3/2$, and the fact that the decaying nature of the superhorizon spectrum during pre-inflation implies $H_{\rm pre-inf}=H_{\rm inf}$, we have
\be
\frac{P_Q^{IR}}{P_Q^{UV}}\sim \left(\frac{k_{IR}}{a H}\right)^{3-2 \nu_{\rm pre-inf}}\gg 1\,,
\ee
leading to an enhancement of power in the IR. This effect is illustrated in Fig. \ref{fig:type_II}.

Clearly the deep IR part of these spectra is unable to accommodate the loss of power hinted in the data. It is however interesting to note that on scales $k\sim a_{60} H_{60}$ there is a region where the oscillating nature of the $\left|C^{(1)}-C^{(2)}\right|$ factor creates a pronounced dip in the power spectrum with a suppression of over 50\% in some cases. It would be interesting to understand if such feature might be sufficient to explain the hints in the data \cite{us:2014}.

\subsection{Pre-inflationary region III}

\begin{figure}[h!]
	\centering
	\begin{minipage}[b]{0.4\linewidth}
	\centering
\includegraphics[width=1\textwidth]{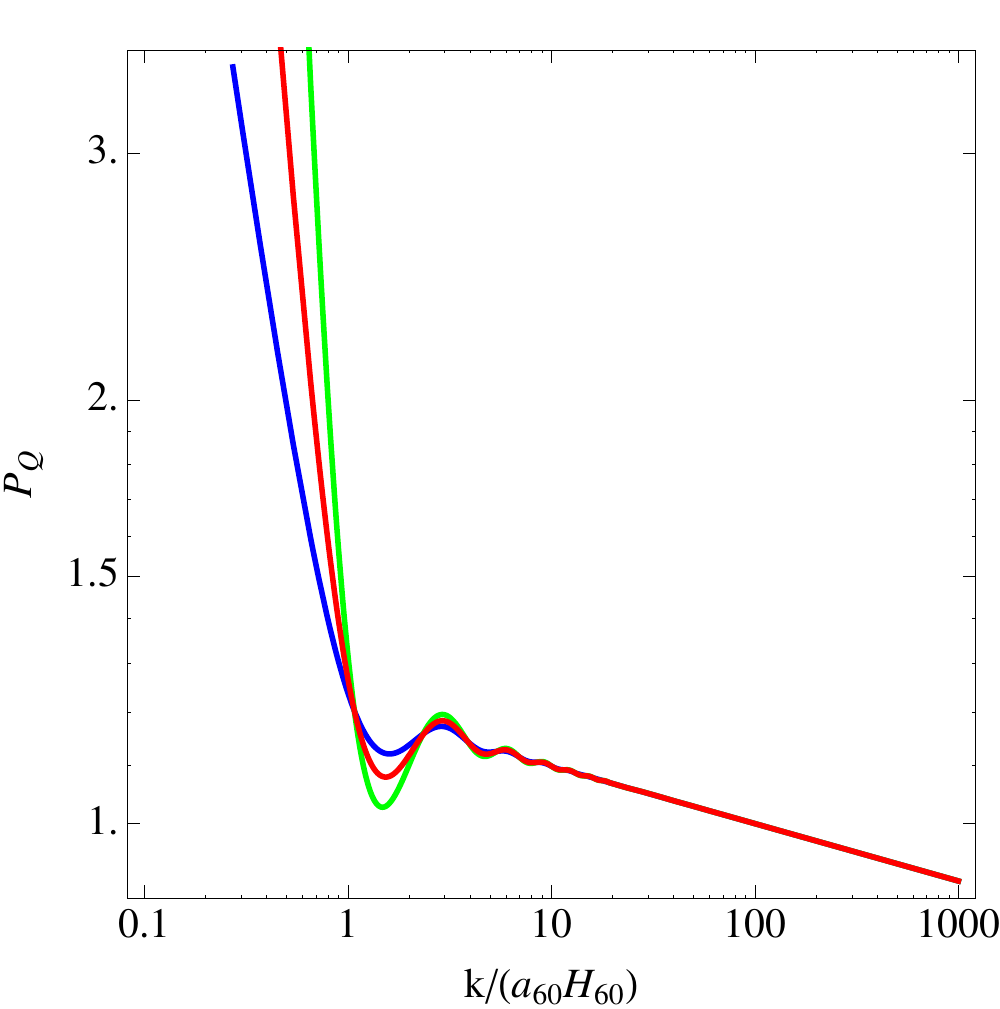}
    \end{minipage}
	\hspace{0.5cm}
	\begin{minipage}[b]{0.41
\linewidth}
	\centering
\includegraphics[width=1\textwidth]{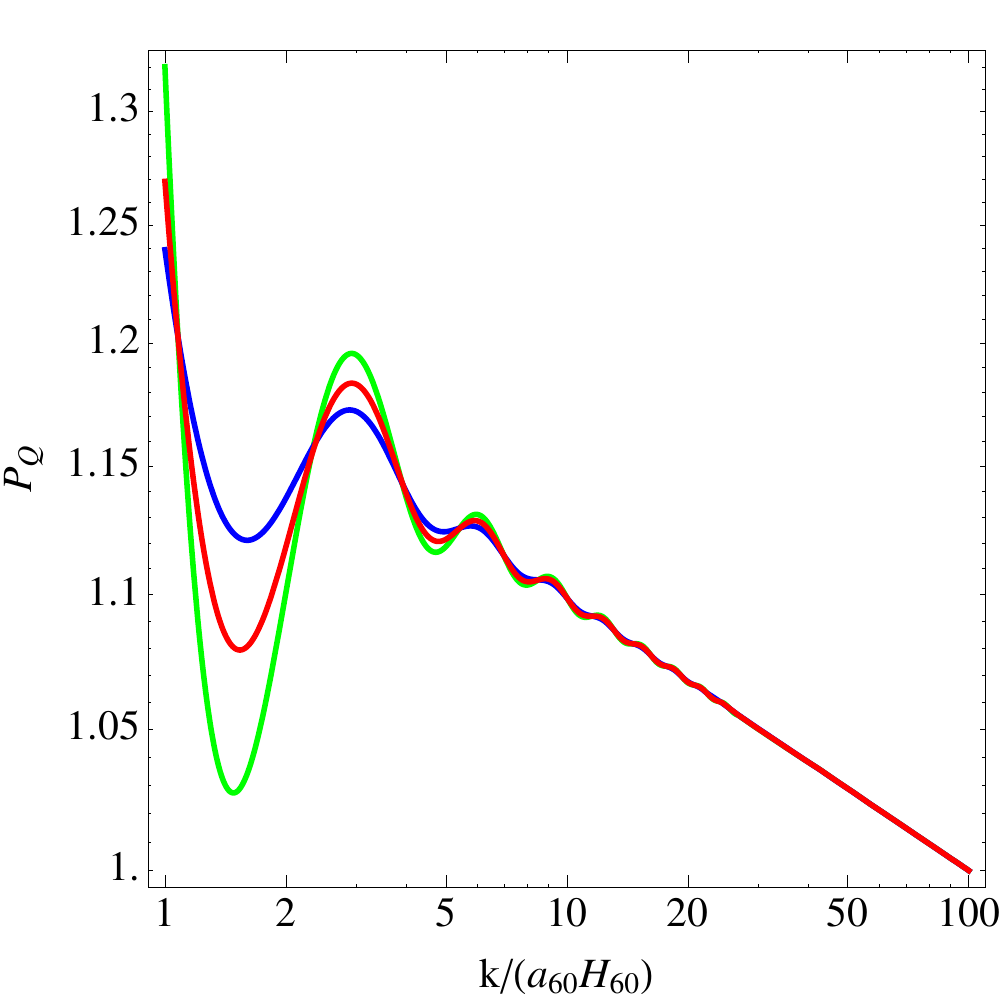}
	\end{minipage}
	\caption{Power spectrum for different type III backgrounds: $\xi=1/2$ (red), $\xi=1/3$ (green) and $\xi=2/3$ (blue). The power on large scales is enhanced due to a red spectral index on those scales.}
	\label{fig:type_III}
\end{figure}

For background solutions of type III the modes that correspond to the low-$\ell$ region of the power spectrum were inside the horizon during the pre-inflationary phase and exited at some point before the transition to slow-roll inflation.
From Eqs. (\ref{PQ1})-(\ref{PQ3}) we find that there is a frozen ($P_Q^1$) and a decaying mode ($P_Q^2+P_Q^3$) in the spectrum and that the frozen mode will dominate in the absence of cancellations in the prefactors. The spectral index for the dominant mode in the pre-inflationary phase is
\be
\left.n_s-1\right|_{\rm pre-inf}\equiv 3-2\nu_{\rm pre-inf}<0\,.
\label{eq:nsIII}
\ee
Given that the spectrum on scales $k\ll a_{60} H_{60}$ does not suffer any modification during the ensuing inflationary phase, we obtain
\be
\frac{P_Q^{IR}}{P_Q^{UV}}\equiv\frac{P_Q^{k\ll a_{60} H_{60}}}{P_Q^{k\gg a_{60} H_{60}}}\sim \left(\frac{k_{IR}}{a H}\right)^{3-2 \nu_{\rm pre-inf}} \gg 1\,.
\ee
Thus we find again a power spectrum that blows-up in the IR due to a red spectral index in Eq. (\ref{eq:nsIII}).
In Fig. \ref{fig:type_III} we plot the primordial spectra for three different type III backgrounds.

\paragraph{Curvature dominated pre-inflation:}
A special situation is a pre-inflationary phase of negative spatial curvature domination with $w=-1/3$ and $\xi=0$ which lies at the boundary between regions II and III. In this case the comoving horizon is constant, implying that there is no horizon-crossing of modes during a curvature dominated epoch. Again the long-wavelength modes which determine the power spectrum at low-$\ell$ are always superhorizon.
The mode functions in a curvature dominated regime are given by growing and decaying exponentials as can be seen from Eq.~\eqref{eq:HankelsCurvature}. The analog of BD initial conditions selects the exponentially decaying mode functions as the initial condition during curvature domination. For this choice, the power spectrum drops to zero for very small $k$.

\subsection{Pre-inflationary region IV}

\begin{figure}[h!]
	\centering
	\begin{minipage}[b]{0.4\linewidth}
	\centering
\includegraphics[width=1\textwidth]{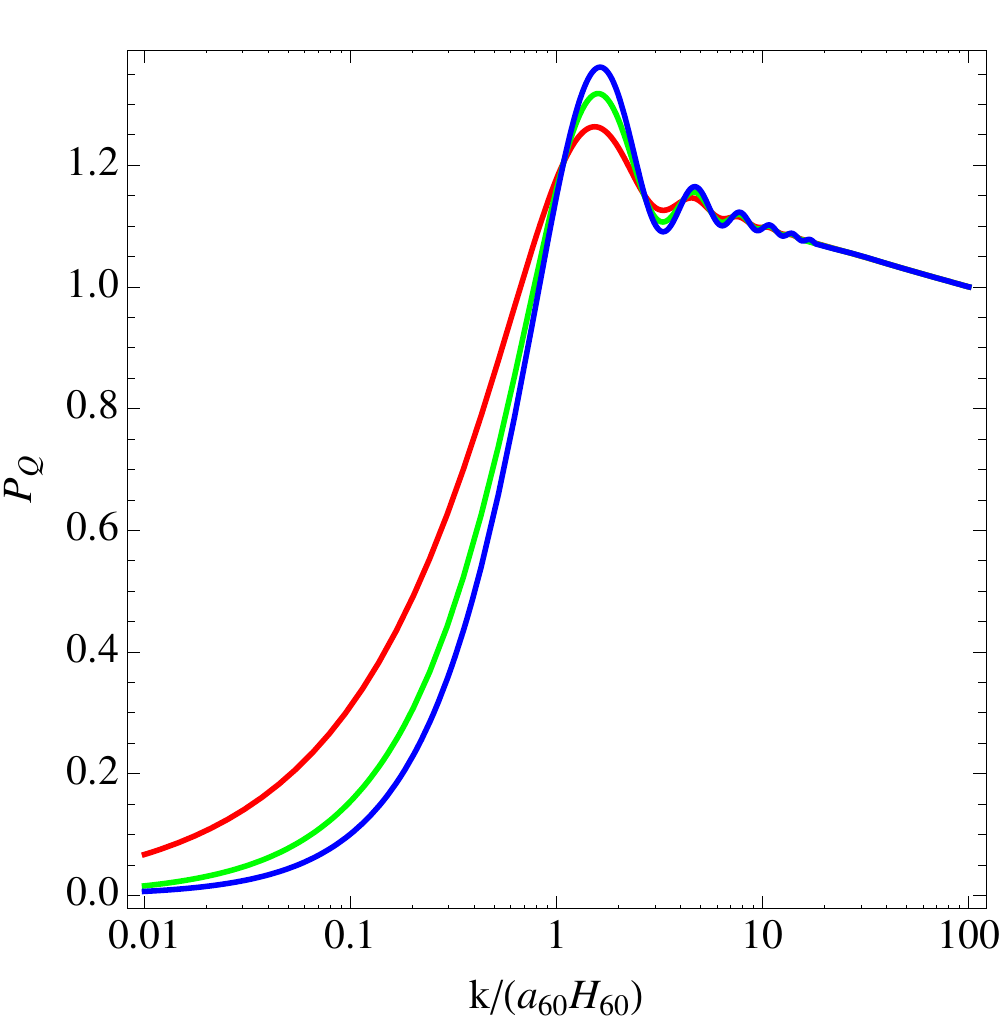}
    \end{minipage}
	\hspace{0.5cm}
	\begin{minipage}[b]{0.415
\linewidth}
	\centering
\includegraphics[width=1\textwidth]{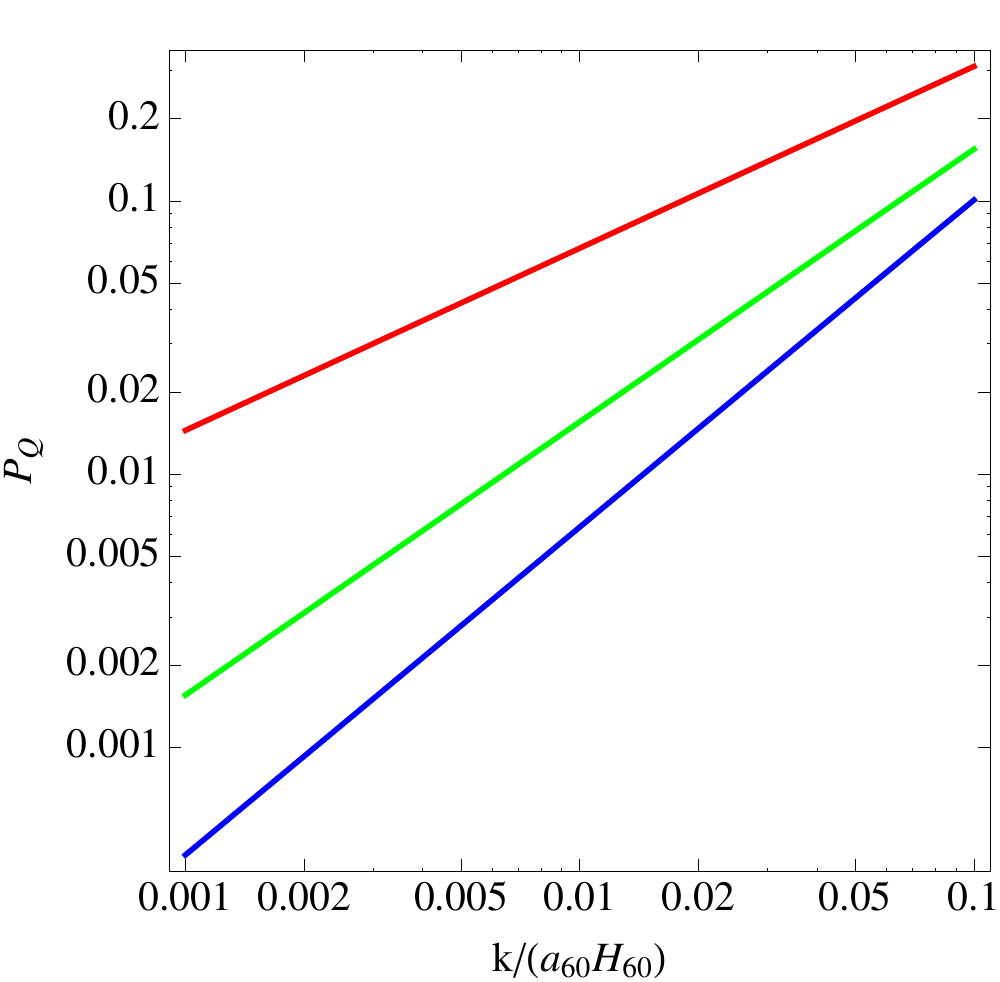}
	\end{minipage}
	\caption{Power spectrum for different type IV backgrounds: $\xi=3/2$ (red), $\xi=2$ (green) and $\xi=5/2$ (blue).
The power on large scales is suppressed due to a blue spectral index on those scales.}
	\label{fig:type_IV}
\end{figure}

For backgrounds of type IV the Hubble parameter $H\propto e^{(\xi-1)N_e}$ grows with $N_e$ and the size of the physical horizon shrinks super-exponentially.
These backgrounds are therefore called super-inflationary solutions \cite{Biswas:2013dry,Liu:2013iha}
and emergent Universe scenarios \cite{Ellis:2002we,Labrana:2013oca}. In these cases, just as for type III backgrounds, the modes corresponding to the large scale part of the power spectrum leave the horizon during the pre-inflationary phase.
The spectrum on large length scales features a dominant frozen blue-tilted part ($P_Q^1$) and a subdominant red-tilted decaying contribution ($P_Q^2+P_Q^3$) given that the spectral index on large scales is
\be
\left. n_s-1\right|_{\rm pre-inf}= 3-2\nu_{\rm pre-inf}>0\,.
\ee
Using the analytic method of Sec. \ref{sec:method} we compute the power spectrum for different type IV backgrounds showing our results in Fig. \ref{fig:type_IV}. Even if the details depend on the precise value of $\xi$, the broad features are the same (similarly to region I): lack of power on the largest scales, oscillatory behaviour on scales close to the horizon at the onset of inflation and a standard red-tilted spectrum of slow-roll inflation on smaller scales.

Let us finally note that a scalar field has an equation of state parameter
\be
w=\frac{p}{\rho}=\frac{{
\cal L}_{kin}-V(\phi)}{{
\cal L}_{kin}+V(\phi)}\,.
\ee
Arranging a suitable evolution of $\phi$ then allows us to provide at least transient regimes with any equation of state parameter $-1<w<1$. This gives us access to regions III and IV even without using exotic sources.

\section{Effects of isotropy and initial conditions}
\label{sec:four}

Having determined the primordial power spectrum from different types of pre-inflationary backgrounds we now undertake a critical assessment of our results.
It is known that modifications of the power spectrum on large scales can be attributed to at least three effects:
\begin{itemize}
\item departures from spatial isotropy,
\item choice of initial conditions for the curvature perturbations,
\item non-slow-roll pre-inflationary dynamics.
\end{itemize}

A complete understanding of the power spectrum on the largest observable scales should include all three effects simultaneously. Should the observational hints evolve into a more significant signal in the future, such a study will be essential to better understand the earliest epochs in the history of the Universe.  In this work we are trying to isolate and understand the effects of the existence of a non-slow-roll pre-inflationary epoch on the power spectrum and therefore make the minimal assumptions about the other two factors: we take spacetime to be isotropic and the curvature perturbations to be in the BD vacuum. While including departures from isotropy in the analysis is rather involved, one can develop some intuition about the effects of non-BD initial conditions.

A departure from the BD vacuum impacts differently on the various solutions discussed in Sec. \ref{sec:pre_inf}. For backgrounds of type I and II (together with the curvature dominated case $\xi=0$) we have seen that the modes on scales $k\ll a_{60} H_{60}$ were always outside the horizon. Regardless of how these modes evolve during pre-inflation, the choice of initial conditions sets the spectral index on these scales and so determines if there is a lack or an excess of power when compared to modes with $k\gg a_{60} H_{60}$ (a similar argument is made in \cite{Powell:2006yg}). The question is then what is the rationale behind the choice of vacuum, and if there is any good reason to move away from minimality and take anything other than BD, like e.g. starting with some excited state. We shall discuss this crucial issue a bit more in Sec. \ref{sec:multiStage} while we wish to mention here that the dependence on the choice of initial conditions is particularly obvious for a curvature-dominated pre-inflationary phase. During curvature domination the usual BD prescription is no longer singled out as the required behavior of small-wavelength modes as there is no horizon-crossing behaviour which evolves modes from deep inside the horizon to superhorizon scales. The mode functions in a curvature dominated regime are given by growing and decaying exponentials as can be seen from Eq.~\eqref{eq:HankelsCurvature}. The analog of BD initial conditions selects the exponentially decaying mode functions as the initial condition during curvature domination. For this choice, the power spectrum drops to zero for very small $k$. However, if there is just curvature domination as the only pre-inflationary epoch, any other choice of initial condition may be used as well, as discussed before.

The results for background solutions of type III and IV are instead more robust, in the sense that modes on scales $k\ll a_{60} H_{60}$ are inside the horizon for some time during pre-inflation and so forget about initial conditions. For these spectra, choosing non-BD initial conditions has little or no impact on the shape of the power spectrum at the end of inflation.

Faced with these observations, we would therefore argue that the more widely considered fast-roll, radiation or matter dominated pre-inflationary phases, being all of type I, do {\it not} by themselves imply a lack of power on large scales. The choice of background merely picks a basis of Hankel functions as solutions to the MS equation. These {\it together} with the choice of vacuum determine the spectral index for small $k$ modes and therefore the scaling of the primordial spectrum with $k$. For pre-inflationary phases where horizon crossing occurs, the very presence of this effect singles out the BD choice of initial conditions relative to the given pre-inflationary phase as the natural boundary conditions in matching the spectrum of local QFT in the UV. Hence the assumption of using natural initial conditions during pre-inflation provides the link between the more widely used type I pre-inflationary phases and power suppression on large angular scales.

\section{Multiple pre-inflationary phases}
\label{sec:multiStage}

The situation described in Sec. \ref{sec:four} can change drastically if we consider multiple pre-inflationary phases. Let us start by considering the effects of a primordial stage of inflation that is followed by a pre-inflationary phase of one of the types described before.

The effects of this primordial inflationary phase on the spectra of type III and IV will not be too significant as they will be more pronounced only on very small scales that are still to enter the horizon today and are therefore unobservable. The same cannot be said for solutions of type I and II, in which the modification of the power spectrum relied on modes that were superhorizon both during pre-inflation and slow-roll.

\begin{figure}[ht!]
 \centerline{\includegraphics[width=0.8\textwidth]{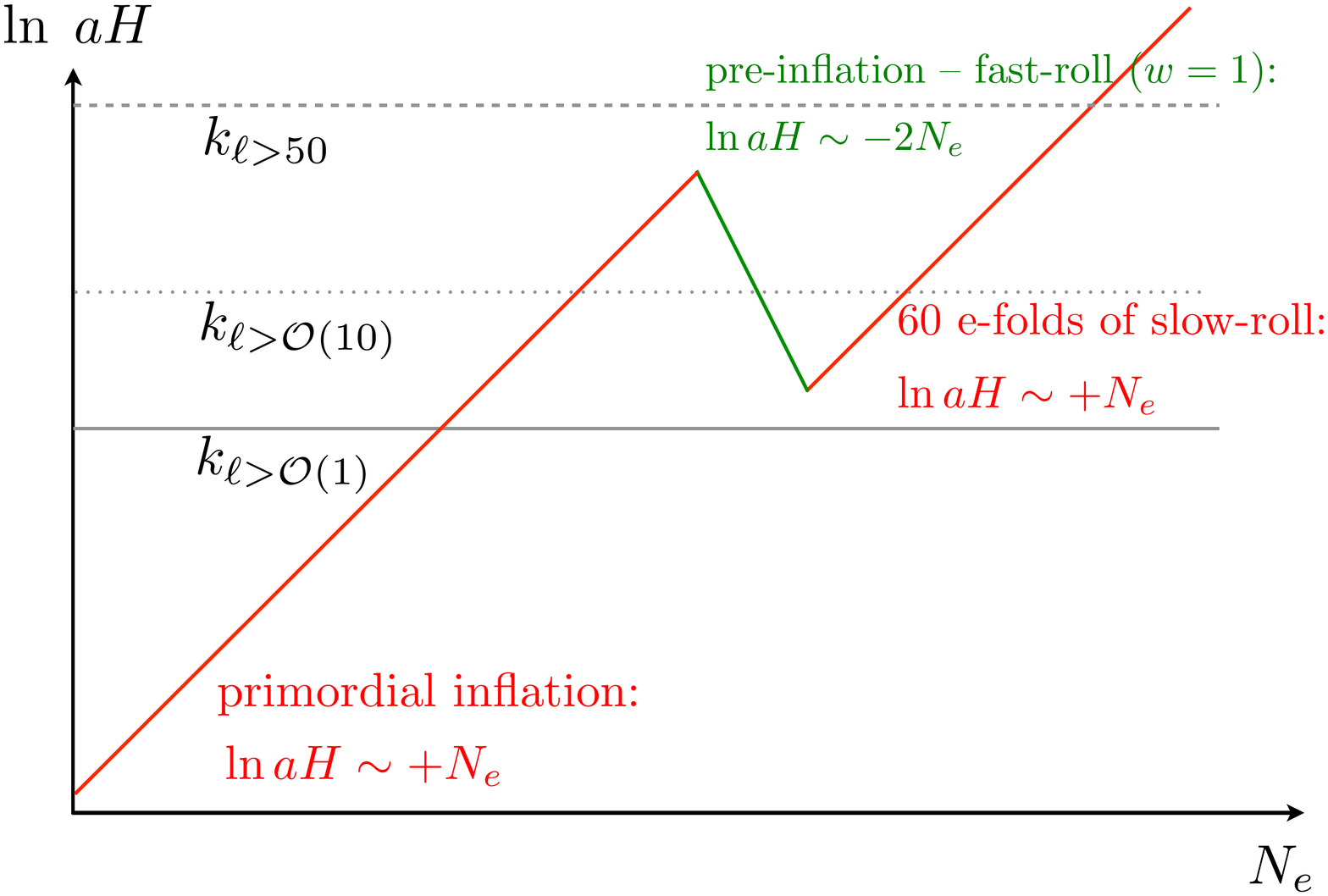}}
 \caption{Logarithmic plot of the comoving horizon $\ln aH$ as a function of $N_e$. The horizontal lines denote fluctuation modes with three different values of the comoving wavenumber $k$. All the modes with $k_{\ell>{\cal O}(1)}$ leave the horizon during the initial stage of inflation and never re-enter during the subsequent pre-inflationary and slow-roll stages. Hence these modes bear memory of the quantum fluctuation of the earliest stage, which here is inflation at $H_{\rm in}\gg H_{\rm fin}$.}\label{fig:5}
\end{figure}

\begin{figure}[h!]
	\centering
	\begin{minipage}[b]{0.4\linewidth}
	\centering
\includegraphics[width=1\textwidth]{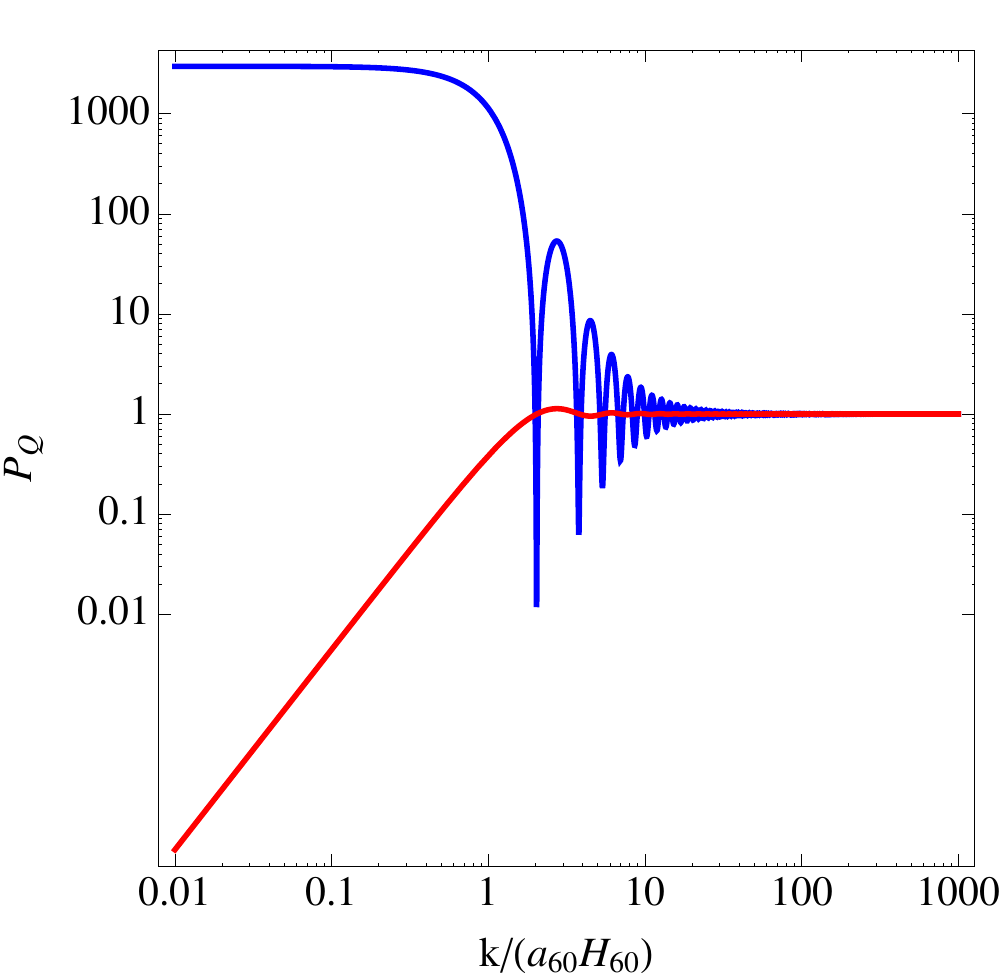}
	\centering
\includegraphics[width=1\textwidth]{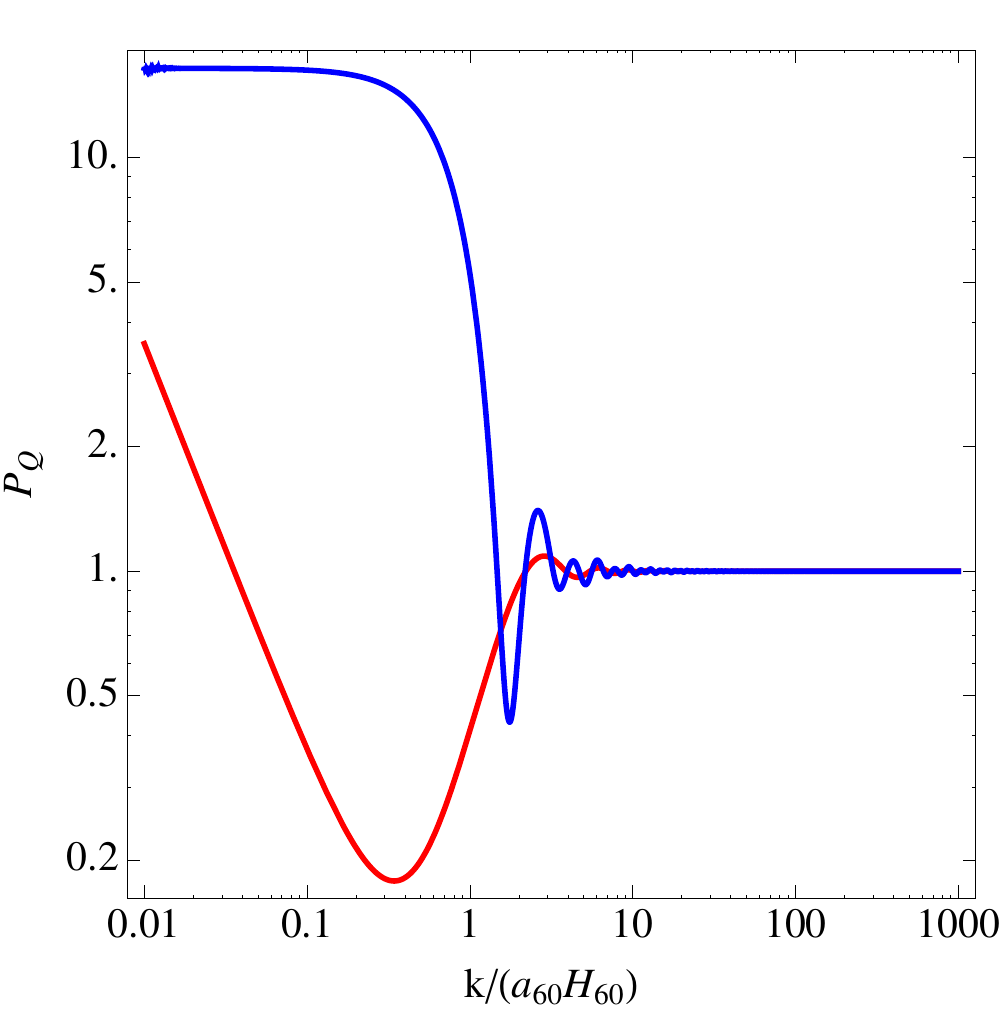}
	\end{minipage}
	\begin{minipage}[b]{0.4
\linewidth}
	\centering
\includegraphics[width=1\textwidth]{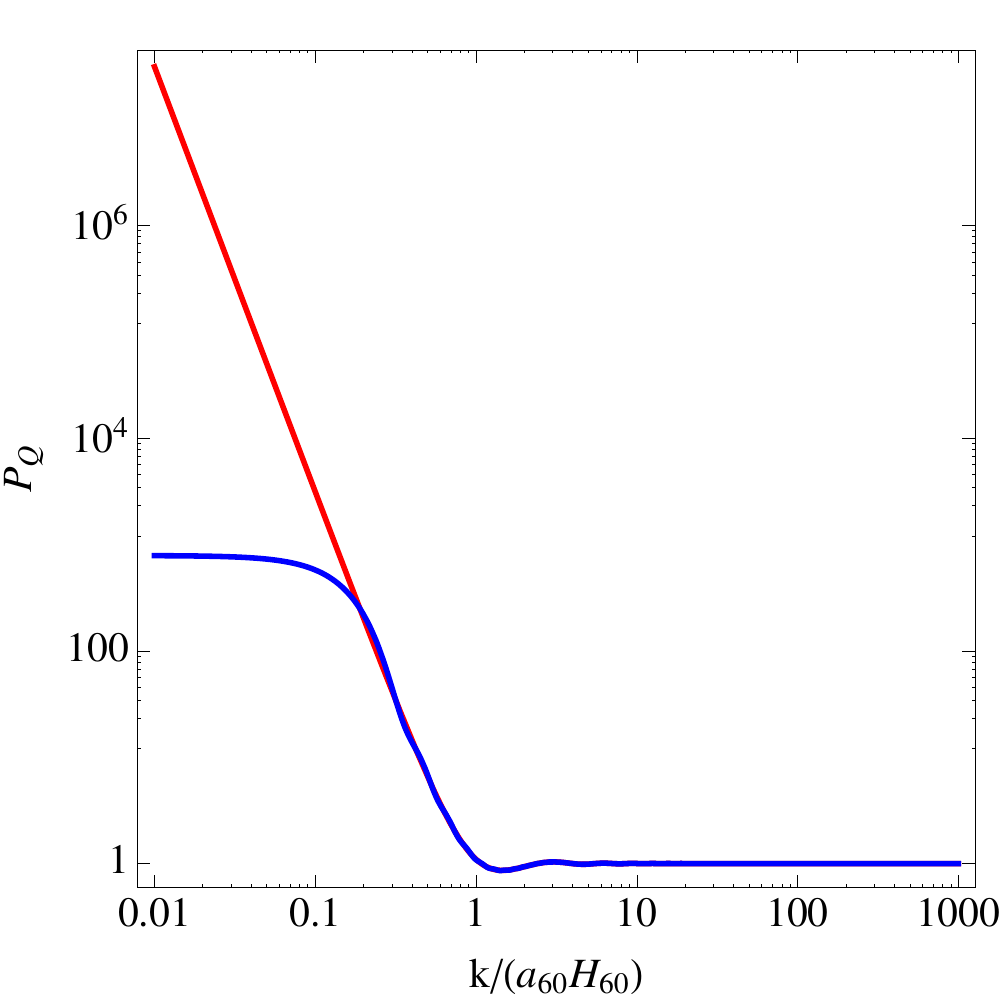}
	\centering
\includegraphics[width=1\textwidth]{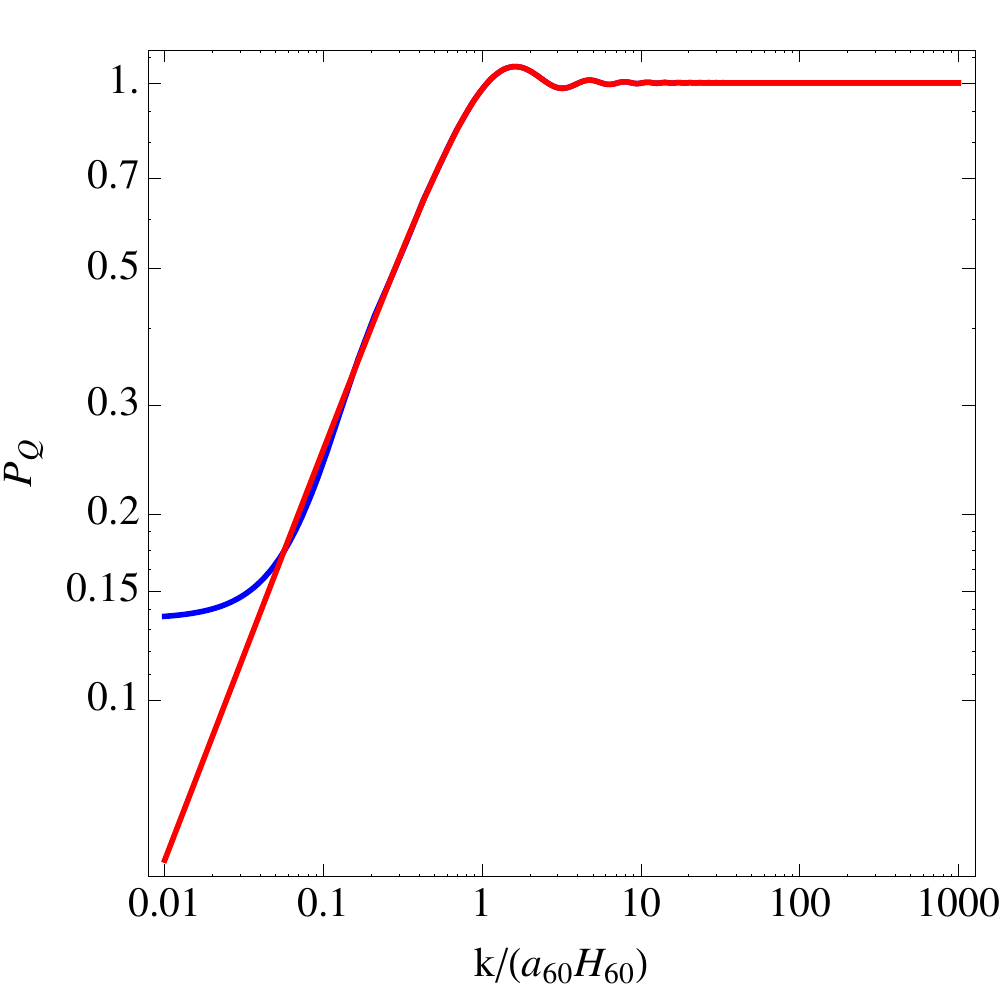}
	\end{minipage}
	\hspace{0.05cm}
	\caption{We plot in blue the effects of a primordial inflationary epoch before type I (top left), II (top right), III (bottom left) and IV (bottom right) pre-inflationary stages. For comparison we plot also in red the spectra obtained assuming a single pre-inflationary phase as in Sec. \ref{sec:pre_inf}.}
	\label{fig:Multiple}
\end{figure}

Looking at Fig.~\ref{fig:5}, we see that modes with $k_{\ell>{\cal O}(1)}$, while superhorizon during pre-inflation and the final slow-roll phase, came from deep inside the horizon during the stage of primordial inflation. These modes now carry memory of that earlier stage of inflation with Hubble rate $H_{\rm in}$ which we expect to be far larger than the Hubble parameter $H_{\rm fin}$ of the final observable stage of inflation: $H_{\rm in} \gg H_{\rm fin}$. Therefore the power spectrum will look like
\be
\label{eq:Pk2PreInf1}
P_Q\sim H_{\rm in}^2\gg H_{\rm fin}^2 \quad\text{for}\quad k\ll a_{60} H_{60}
\qquad\text{and}\qquad
P_Q\sim H_{\rm fin}^2 \quad\text{for}\quad k\gg a_{60} H_{60}\,.
\ee
We expect the intermediate regime $k\eta \simeq 1$ to show strong oscillatory patterns from interference terms.
Because of an initial inflationary stage, large scale modes loose memory of initial conditions but now, as can be seen from Eq. (\ref{eq:Pk2PreInf1}),
the power spectrum at small $k$ is enhanced instead of being suppressed as in the case I without prior inflation.
This qualitative behaviour is confirmed by an analytical computation whose results are displayed in Fig. \ref{fig:Multiple}.

\paragraph{False-vacuum inflation prior to curvature dominated pre-inflation:} Let us make some comments on the special case of a curvature dominated pre-inflationary phase that arises as a tunneling event out of false dS vacuum producing a bubble internally dominated by negative spatial curvature. Choosing only the exponentially decaying modes as the initial condition during the curvature dominated post-tunneling epoch with their vanishing power spectrum at $k\to 0$ is now consistent with the fact that in a pure false vacuum dS phase there is no curvature perturbation~\cite{Linde:1999wv,Yamauchi:2011qq}. This is a particular case exemplifying the discussion at the end of Sec. ~\ref{sec:four} -- namely that a particular choice of initial conditions during the last epoch of pre-inflation immediately prior to the onset of slow-roll, is responsible for the vanishing of the curvature perturbation at very small $k$, and this particular choice is generated dynamically here by the presence of an earlier false-vacuum dS pre-inflationary epoch.

\begin{figure}[t!]
\centerline{\includegraphics[width=0.5\textwidth]{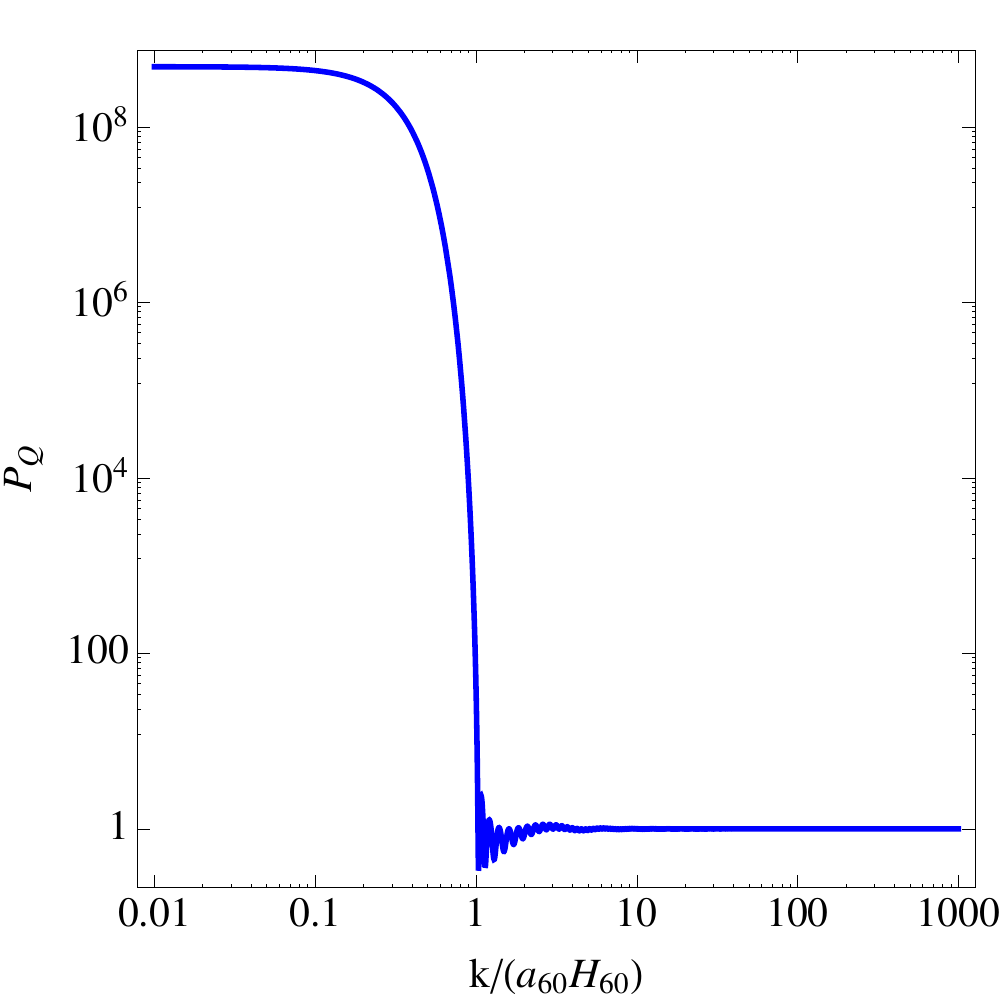}}
\caption{The power spectrum $P_Q$ which is proportional to the tensor mode spectrum of a 3-epoch evolutionary sequence: starting in a false-vacuum dS epoch, we tunnel into an epoch of curvature domination followed by the final observable $60$ e-foldings of slow-roll.}\label{fig:dScurvSlowRoll}
\end{figure}

On the other hand there are tensor perturbations created during false-vacuum dS inflation whose power spectrum is controlled by $P_Q$ since $P_T\sim P_Q\sim H_{\rm in}^2 \gg H_{\rm fin}^2$. Therefore, we find a strongly enhanced flat tensor power spectrum at very small $k$, while at $k\simeq a_{60}H_{60}$ we will find strong oscillations with increasing amplitude towards smaller $k$ with no suppression (see Fig.~\ref{fig:dScurvSlowRoll}).
Tensor perturbations convert at large angular scales $\ell < 50$ into a contribution $(\delta C^{TT}_\ell)_{\rm tensor}$ to the CMB temperature two-point function. Hence, the strongly rising tensor power spectrum of a two-stage pre-inflationary epoch consisting of false-vacuum dS inflation followed by tunneling-induced curvature domination will give rise to a rapidly increasing $C^{TT}_\ell$-power spectrum at low-$\ell$ as well.

Note that this behaviour is consistent with similar results in~\cite{Yamauchi:2011qq}, in particular after combining the effects of both curvature perturbations and tensor modes on the temperature $C_\ell$'s. The ensuing power spectrum of this false-vacuum dS $\to$ curvature domination $\to$ slow-roll sequence hence looks qualitatively similar to the case of a primordial epoch of slow-roll inflation preceding a fast-roll pre-inflationary phase
(see upper left plot of Fig.~\ref{fig:type_III}).

This general result -- power suppression at low-$\ell$ if there is no earlier phase of inflation or curvature domination, and power enhancement at low-$\ell$ if there is earlier inflation or curvature domination -- has direct implications for the string landscape. The string landscape inevitably has either false-vacuum or slow-roll inflation at a much higher scale of $H$ to the past of our own observable slow-roll phase. As the exit from that primordial stage of inflation proceeds either by fast-roll or some form of tunneling followed by fast-roll, the detection of a power suppression (not just a dip as in regions II or III) at low-$\ell$ would pose a severe challenge to the string landscape.\footnote{In particular because tunneling vacuum transitions could produce bubbles with negative internal spatial curvature. While this is true for Coleman-de Luccia transitions, this is an open question for brane-flux transitions of Brown-Teitelboim type \cite{Brown:1988kg,deAlwis:2013gka} which dominate the type IIB 3-form flux vacuum landscape.}

\section{Discussion and conclusions}
\label{sec:discuss}

In this paper we showed that models where slow-roll inflation lasted `just enough' to solve the flatness and horizon problems,
i.e. about $50-60$ e-foldings but not more, necessarily feature modifications of the CMB power spectrum at large angular scales.
These low-$\ell$ modes left the horizon just before the beginning of these $50-60$ e-foldings of inflation or
were never inside the horizon but with a scale slightly larger than the horizon size at the beginning of slow-roll inflation.

We performed a systematic and model-independent analysis of any possible non-slow-roll background evolution prior to the final stage of slow-roll inflation.
Assuming an isotropic spacetime and BD initial conditions, we found that a lack or an excess of power at large scales depends just on the value of the equation of state parameter. We found a high degree of universality since most common backgrounds like fast-roll evolution, matter or radiation-dominance give rise to a power loss at low-$\ell$ with very similar features: a peak together with an oscillatory behaviour at scales around the value of the Hubble parameter at the beginning of slow-roll inflation and a suppression of power in the IR. This implies that by seeing a peak in the CMB spectrum we would indirectly see the beginning of inflation!

An important observation to trust these results is that the presence of a lack or an excess of power at low-$\ell$ depends deeply on the choice of initial conditions. This choice is in practice irrelevant for modes which were inside the horizon in the past but becomes crucial for low-$\ell$ modes which were always superhorizon. Given that this is the case for most of the common backgrounds mentioned above, our results apply only to the case of BD initial conditions.
A fundamental question to ask is then: what sets the initial conditions for modes which were never in causal contact with us?

We tried to answer this question by studying scenarios with multiple pre-inflationary stages with an initial primordial inflationary epoch with Hubble parameter $H_{\rm in}$, followed by a non-slow-roll phase and finally by a standard inflationary era with Hubble parameter $H_{\rm fin}$ which lasted only $50-60$ e-foldings. Thanks to this primordial stage of inflation, these low-$\ell$ modes were inside the horizon in the past, and so any dependence on initial conditions gets removed. However the price to pay is that this new scenario gives rise to an excess, instead of a lack, of power at large scales since the amplitude of the IR spectrum is set by $H_{\rm in}$ which we expect to be larger than $H_{\rm fin}$ which set the amplitude of the spectrum in the UV.

However there have been mounting observational hints of a CMB power loss at large angular scales from COBE, WMAP and Planck.
Planck found a power deficit of $5-10\%$ at $\ell \lesssim 40$ with a statistical significance of $2.5-3\sigma$ which would increase to $3.5-4\sigma$
in the presence of large tensor modes that might have been recently detected by BICEP2.
If not a statistical fluke, this power loss could be the signal that inflation lasted `just enough' and was precedented by a non-slow-roll epoch.
Assuming BD initial conditions, a pre-inflationary era characterised by fast-rolling, matter or radiation dominance would yield exactly this kind of behaviour in the absence of a primordial inflationary stage. Moreover, we have seen that a power loss is obtained also in the case of non-slow-roll backgrounds with $\xi>1$ where $H$ increases giving rise to a super-accelerated expansion like in the case of super-inflation ($\xi=2$).
On the other hand, more exotic backgrounds corresponding to regions II and III would naively be disfavoured since they lead to an enhancement of power at large scales.

\begin{figure}[t!]
 \centering
	\includegraphics[width=0.8\textwidth]{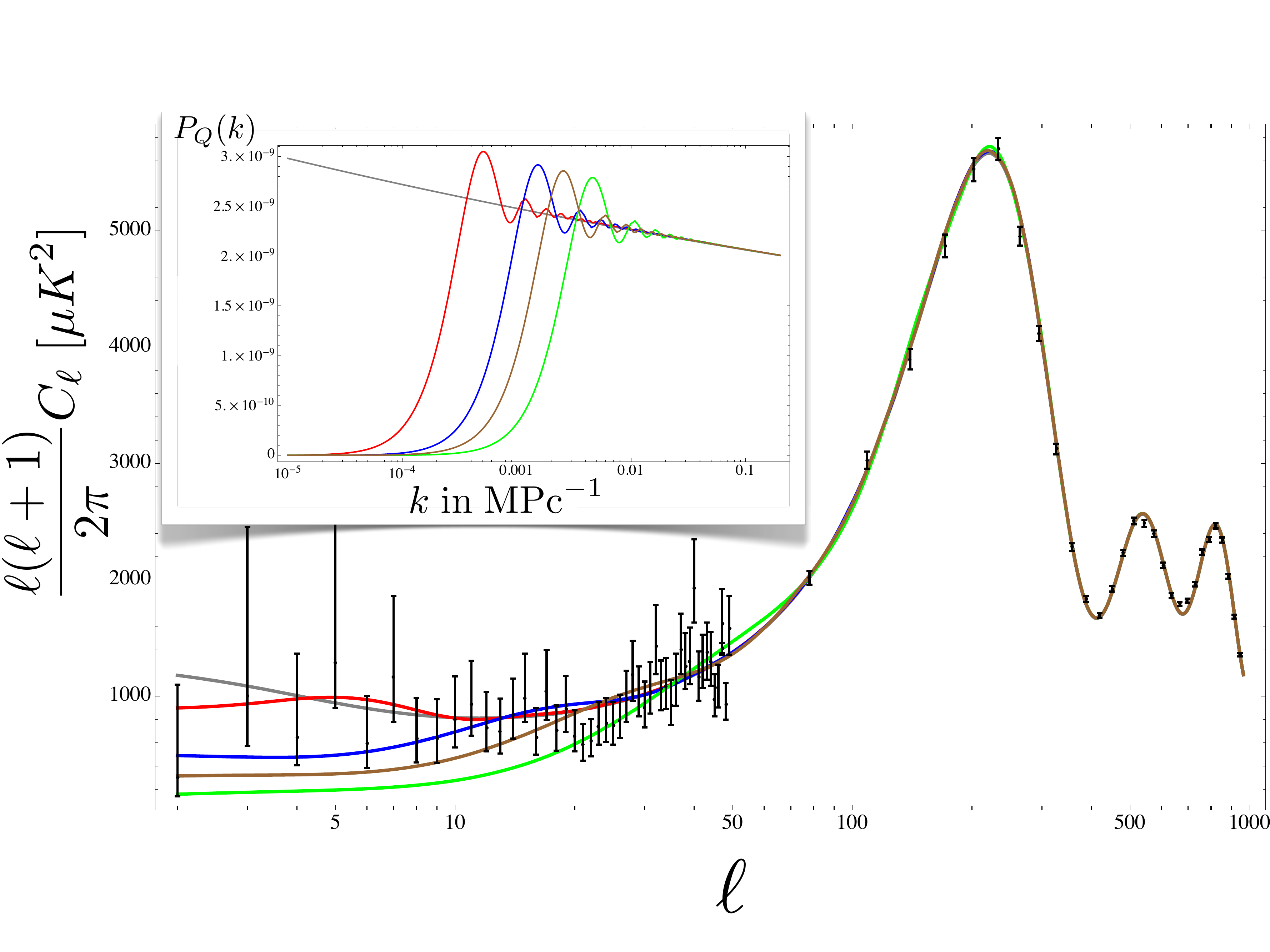}
	\vspace*{-5ex}
 \caption{$C_\ell$ power spectrum for a region I fast-roll pre-inflationary stage with $\xi=-2$ ($w=1$, $\nu=0$) compared with the Planck 2013 CMB power spectrum. Different colours correspond to different positions of the cut-off in the power spectrum.}\label{fig:ClregI}
 \vspace*{-2ex}
\end{figure}

\begin{figure}[t!]
 \centering
	\includegraphics[width=0.8\textwidth]{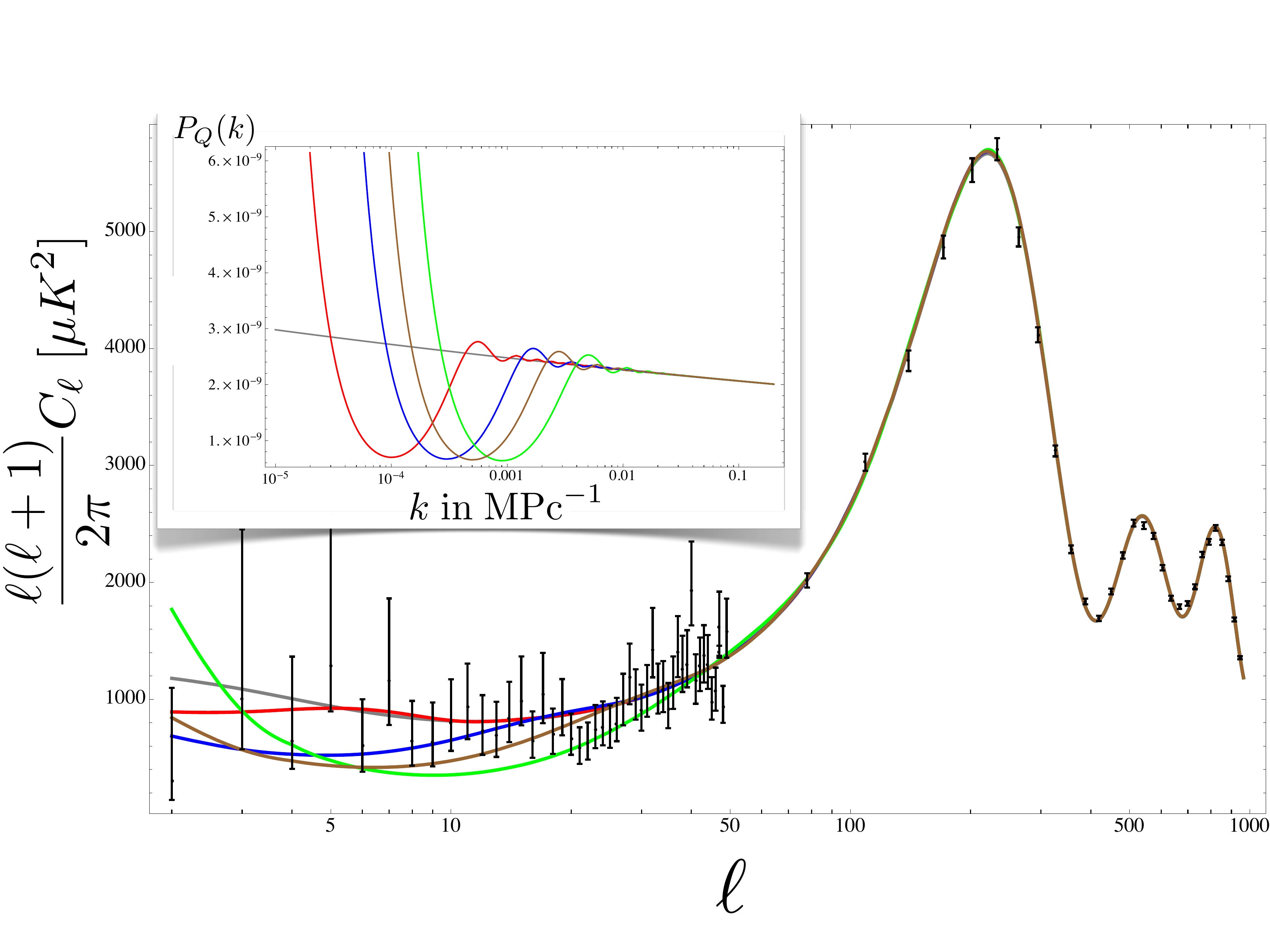}
	\vspace*{-5ex}
 \caption{$C_\ell$ power spectrum for a region II pre-inflationary stage with $\xi=-1/3$ ($w=-1/9$, $\nu=5/2$) compared with the Planck 2013 CMB power spectrum. Different colours correspond to different positions of the cut-off in the power spectrum.}\label{fig:ClregII}
\end{figure}

A detailed fit to Planck data is beyond the scope of this paper, and so we leave it for the future \cite{us:2014}. However, in Figs. \ref{fig:ClregI} and \ref{fig:ClregII} we present two illustrative $C_\ell$-plots, comparing a region I and a region II power spectrum with Planck 2013 data. Different colours correspond to different positions of the cut-off in the power spectrum. The smaller the cut-off, the longer inflation would have lasted and the less noticeable the effects of pre-inflation would be.

Let us finally conclude by stressing that `just enough' inflation would be a stunning opportunity to test physics before inflation by seeing modifications of the power spectrum at low-$\ell$ but would also open up a `coincidence problem' that would deserve a deep explanation: why inflation lasted just $50-60$ e-foldings and not more? Both in the string landscape \cite{Freivogel:2005vv} and in particular string inflationary models \cite{Pedro:2013pba,Cicoli:2013oba} there is a statistical tendency towards small numbers of e-foldings but we believe that this crucial question would deserve a more thorough explanation.

\section*{Acknowledgments}

We are grateful to Thorsten Battefeld, Fabio Finelli, Marco Peloso and Alexei Starobinsky for illuminating discussions and comments.
This work was supported in parts by  DOE Grant Nos. DE-FG02-13ER42020 and Impuls und Vernetzungsfond of the Helmholtz Association of German Research Centres under grant HZ-NG-603, and German Science Foundation (DFG) within the Collaborative Research Center 676 ``Particles, Strings and the Early Universe''.

\appendix

\section{Recursive relations for the power spectrum}
\label{App}

The simplifying assumptions about the background explained in the main text, together with the requirement of continuity of the function $u(N_e)$ and of its first derivative across each transition in the history of the Universe, allows one to recursively compute the power spectrum given a certain background and choice of initial conditions for the perturbations.

Continuity across the boundary separating the $i+1$ from the $i$-th phase allows us to express the `new' integration constants in terms of the `old' ones  as
\be
\left(
\begin{array}{c}
C_{i+1}^{(1)}\\
C_{i+1}^{(2)}\\
  \end{array} \right)=
\mathcal{A} \times \left(
\begin{array}{c}
C_{i}^{(1)}\\
C_{i}^{(2)}\\
  \end{array} \right),
\ee
where $\mathcal{A}$ is a matrix whose entries are written in terms of products of Hankel functions as
(defining $r\equiv \frac{k}{a H }\big|_{N_{e,i}}$)

\begin{eqnarray}
\mathcal{A}_{1,1}&\equiv& \frac{i\pi}{8} \frac{r}{\sqrt{\xi_i \xi_{i+1}}}
\left\{\left[ H_{\nu_i+1}^{(1)}\left(\frac{r}{\xi_i}\right)
-H_{\nu_i-1}^{(1)}\left(\frac{r}{\xi_i}\right)\right] H_{\nu_{i+1}}^{(2)}\left(\frac{r}{\xi_{i+1}}\right)\right. \\
&+& \left. H_{\nu _i}^{(1)}\left(\frac{r}{\xi _i}\right) \left[H_{\nu_{i+1}-1}^{(2)}\left(\frac{r}{\xi _{i+1}}\right)
-H_{\nu_{i+1}+1}^{(2)}\left(\frac{r}{\xi _{i+1}}\right)\right]\right\} -\frac{i \pi}{8} \frac{\xi_i-\xi_{i+1}}{\sqrt{\xi _i \xi _{i+1}}} H_{\nu_i}^{(1)}\left(\frac{r}{\xi _i}\right) H_{\nu_{i+1}}^{(2)}\left(\frac{r}{\xi_{i+1}}\right), \nonumber
\end{eqnarray}
\begin{eqnarray}
\mathcal{A}_{1,2}&\equiv& \frac{i \pi}{8} \frac{r}{\sqrt{\xi _i \xi _{i+1}}}
\left\{\left[H_{\nu_i+1}^{(2)}\left(\frac{r}{\xi _i}\right)-H_{\nu _i-1}^{(2)}\left(\frac{r}{\xi_i}\right)\right]
H_{\nu _{i+1}}^{(2)}\left(\frac{r}{\xi_{i+1}}\right)\right. \\
&+& \left. H_{\nu _i}^{(2)}\left(\frac{r}{\xi _i}\right) \left[H_{\nu _{i+1}-1}^{(2)}\left(\frac{r}{\xi _{i+1}}\right)
-H_{\nu_{i+1}+1}^{(2)}\left(\frac{r}{\xi _{i+1}}\right)\right]\right\} -\frac{i \pi}{8} \frac{\xi _i-\xi _{i+1}}{\sqrt{\xi _i \xi_{i+1}}}
H_{\nu _i}^{(2)}\left(\frac{r}{\xi _i}\right) H_{\nu_{i+1}}^{(2)}\left(\frac{r}{\xi _{i+1}}\right), \nonumber
\end{eqnarray}
\begin{eqnarray}
\mathcal{A}_{2,1}&\equiv&\frac{i\pi}{8} \frac{r}{\sqrt{\xi _i \xi _{i+1}}}
\left\{\left[H_{\nu_{i+1}+1}^{(1)}\left(\frac{r}{\xi _{i+1}}\right)-H_{\nu_{i+1}-1}^{(1)}\left(\frac{r}{\xi _{i+1}}\right)\right]
H_{\nu _i}^{(2)}\left(\frac{r}{\xi _i}\right)\right. \\
&+& \left. H_{\nu _{i+1}}^{(1)}\left(\frac{r}{\xi_{i+1}}\right) \left[H_{\nu _i-1}^{(2)}\left(\frac{r}{\xi _i}\right)
-H_{\nu_i+1}^{(2)}\left(\frac{r}{\xi _i}\right)\right]\right\} + \frac{i\pi}{8} \frac{\xi _i-\xi _{i+1}}{\sqrt{\xi _i \xi_{i+1}}}
H_{\nu _{i+1}}^{(1)}\left(\frac{r}{\xi _{i+1}}\right) H_{\nu_i}^{(2)}\left(\frac{r}{\xi _i}\right), \nonumber
\end{eqnarray}
\begin{eqnarray}
\mathcal{A}_{2,2}&\equiv&\frac{i\pi}{8} \frac{r}{\sqrt{\xi _i \xi _{i+1}}}
\left\{\left[H_{\nu_i-1}^{(1)}\left(\frac{r}{\xi _i}\right)-H_{\nu _i+1}^{(1)}\left(\frac{r}{\xi_i}\right)\right]
H_{\nu _{i+1}}^{(1)}\left(\frac{r}{\xi_{i+1}}\right)\right. \\
&+& \left. H_{\nu _i}^{(1)}\left(\frac{r}{\xi _i}\right) \left[H_{\nu _{i+1}+1}^{(1)}\left(\frac{r}{\xi _{i+1}}\right)
-H_{\nu_{i+1}-1}^{(1)}\left(\frac{r}{\xi _{i+1}}\right)\right]\right\}+\frac{i\pi}{8} \frac{\xi _i-\xi _{i+1}}{\sqrt{\xi _i \xi_{i+1}}}
H_{\nu _i}^{(1)}\left(\frac{r}{\xi _i}\right) H_{\nu_{i+1}}^{(1)}\left(\frac{r}{\xi_{i+1}}\right). \nonumber
\end{eqnarray}
When considering a multi-stage history, all one then needs to do is to take the suitable product of $\mathcal{A}$ matrices to find the post-inflationary power spectrum
\be
\left(
\begin{array}{c}
C_{i_{\rm max}}^{(1)}\\
C_{i_{\rm max}}^{(2)}\\
  \end{array} \right)=
\mathcal{A}^{i_{\rm max}\rightarrow i_{\rm max}-1}\times...\times\mathcal{A}^{3\rightarrow2}\times \mathcal{A}^{2\rightarrow1} \times \left(
\begin{array}{c}
C_{1}^{(1)}\\
C_{1}^{(2)}\\
  \end{array} \right)
\ee
where $C_{1}^{(1)}$ and $C_{1}^{(2)}$ are determined by the choice of initial conditions for the mode functions.

We must stress that even though the same method applies for histories involving curvature dominated phases ($\xi=0$), the above expressions hold only for Universes in which $\xi\neq 0$ throughout.

\bibliographystyle{JHEP.bst}
\bibliography{low-l-v10}
\end{document}